\documentstyle[12pt,a4,amstex,amscd,axodraw,epsfig]{article}
\newcommand\slv{v\kern-5pt\raise1pt\hbox{$\scriptstyle/$}\kern1pt}

\newcommand\be{\begin{equation}}
\newcommand\ee{\end{equation}}
\newcommand{\eps}{\varepsilon}
\def\bq{\begin{eqnarray}}
\def\eq{\end{eqnarray}}

\begin{document}
\thispagestyle{empty}
\begin{flushright}
NIKHEF-99-010\\
\end{flushright}
\vspace{0.5cm}
\begin{center}
{\Large \bf Equivariant dimensional regularization}\\[.3cm]
\vspace{1.7cm}
$\mbox{{\sc \bf Stefan Weinzierl}}^\ast$ \\[1cm]
\begin{center} \em 
NIKHEF, P.O. Box 41882, NL - 1009 DB Amsterdam, The Netherlands
\end{center}\end{center}
\vspace{2cm}

\begin{abstract}\noindent
{The calculation of loop amplitudes with parity violation or spin effects within dimensional
regularization needs a consistent definition of $\gamma_5$. 
Also loop calculations in supersymmetric theories need a consistent definition of $\gamma_5$.
In this paper we develop a new formalism, which allows us to define consistent
regularization schemes.
We use Grothendieck's $K$-functor to construct finite-dimensional vectorspaces of non-integer rank.
The rank will play the r\^ole of the ``$4-2\eps$'' in conventional dimensional regularization.
We then define two regularization schemes, one similar to the 't Hooft--Veltman scheme, the other
one as a scheme, where all algebra is performed in four dimensions.
Lorentz invariance is maintained in both cases. However the structure of the Clifford algebra
cannot be preserved. We show that the HV-like scheme and the four-dimensional scheme correspond
to two different deformations of the Clifford algebra.
It is the purpose of this paper to advocate the four-dimensional scheme for future calculations,
since it is easier to use.
As a consistency check we performed explicit one-loop calculations of various triangle anomalies 
in both schemes 
and we found agreement with Bardeen's results.
}
\end{abstract}

\vspace*{\fill}

\noindent 
$^\ast${\small email address : stefanw@@nikhef.nl}

\newpage

\section{Introduction}

Multi-loop calculations need a regularization of the loop-momentum integrals in order to keep
track of infinities and are most conveniently performed within the framework
of dimensional regularization \cite{HV,Bollini,Ashmore}.
For an introduction to dimensional regularization, see the review by Leibbrandt \cite{Leibbrandt}
or the book by Collins \cite{CDR}.
As long as one deals only with non-chiral and non-supersymmetric theories and if one is only
interested in unpolarized quantities, dimensional regularization has the nice
feature that it preserves the BRS-symmetry.
However, this statement is no longer true as soon as one considers
loop amplitudes with parity violation or spin-effects within dimensional regularization.
These calculations require the definition of $\gamma_5$ in $d$ dimensions.
Another example where a consistent definition of $\gamma_5$ is required is given by supersymmetric
theories.
The algebraic properties which $\gamma_5$ has in four dimensions
($\gamma^2_5=1$, $\{\gamma_\mu, \gamma_5 \} = 0$, 
$\mbox{Tr}\; \gamma_\mu \gamma_\nu \gamma_\rho \gamma_\sigma \gamma_5 = 4 i \eps_{\mu\nu\rho\sigma}$)
cannot be maintained simultaneously within dimensional regularization.
A naive approach, which maintains the anticommuting property of $\gamma_5$ in $d$
dimensions, is inconsistent. The difficulties are related to Dirac traces like
\bq
\eps^{\mu\nu\rho\sigma} \mbox{Tr}\;\left(\gamma_\tau \gamma_\mu \gamma_\nu \gamma_\rho \gamma_\sigma \gamma^\tau \gamma_5 \right),
\eq
where $\eps_{\mu\nu\rho\sigma}$ equals 1, if $(\mu\nu\rho\sigma)$ is an
even permutation of $(0123)$, equals -1, if $(\mu\nu\rho\sigma)$ is an
odd permutation of $(0123)$ and equals 0 otherwise.
Using the cyclicity of the trace and the anticommuting relations of $\gamma_5$
one derives
\bq
\eps^{\mu\nu\rho\sigma} \left( d - 4 \right) \mbox{Tr}\;\left(\gamma_\mu \gamma_\nu \gamma_\rho \gamma_\sigma \gamma_5 \right) & = & 0.
\eq
At $d=4$ this equation permits the usual non-zero trace of $\gamma_5$ with four
other Dirac matrices. However, for $d \neq 4$ we conclude
that the trace equals zero, and there is no smooth limit $d \rightarrow 4$ which 
reproduces the non-zero trace at $d=4$.
Despite this inconsistency it is generally believed that the naive approach will give correct
results for a fermion loop with an even number of $\gamma_5$'s inserted \cite{naive1,naive2,naive3}.
T.L. Trueman \cite{Trueman} gave a set of rules, derived from the 't Hooft-Veltman scheme, which in
some cases reduces to the naive prescription.\\
\\
The general accepted scheme is the one originally proposed by 
't Hooft and Veltman \cite{HV} and by Akyeampong and Delbourgo \cite{Akyeampong}
and systematized by Breitenlohner and Maison \cite{Breitenlohner}.
It was furhter considered by Thompson and Yu \cite{Thompson} and Collins \cite{CDR}. \\
Here $\gamma_5$ is defined as a generic four dimensional object:
\bq
\gamma_5 & = & \frac{i}{4!} \eps_{\mu\nu\rho\sigma} \gamma^\mu \gamma^\nu 
\gamma^\rho \gamma^\sigma.
\eq
In the 't Hooft-Veltman scheme $\gamma_5$ anticommutes with the first four Dirac matrices, but commutes with the remaining ones:
\bq
\left\{ \gamma^\mu , \gamma_5 \right\} = 0, & & \mbox{if} \;\; \mu=0,1,2,3, \\
\left[ \gamma^\mu, \gamma_5 \right] = 0 & & \mbox{otherwise}.
\eq
This forces us to distinguish carefully between $d$-dimensional quantities and
four-dimen\-sional quantities. Calculations in the 't Hooft--Veltman scheme are
therefore very elaborate.
The fact that $\gamma_5$ no longer anticommutes with all Dirac matrices 
gives rise to physical and non-physical anomalies.
Physical anomalies, like the triangle anomaly \cite{Adler,Bell,Bardeen,Adler2}, are for example
of relevance for the decay of a pion into two photons.
In the standard model physical anomalies cancel if the sum over all fermion species is performed.
Non-physical anomalies are an artefact of the regularization or renormalization scheme.
They can (and have to) be removed by appropriate chosen counterterms \cite{Bonneau,Larin,Martin}.
Since the 't Hooft-Veltman scheme does not respect supersymmetry, it is usually not used for calculations
in supersymmetric theories.
Finally we note that there are two approaches for a rigourous definition of the 't Hooft--Veltman
scheme:
The constructive approach of Wilson \cite{Wilson} and Collins \cite{CDR} takes the four-dimensional
Minkowski space as a subspace of an infinit-dimensional space. All objects, like Dirac matrices
and the metric tensor, are explicitly given in that infinit-dimensional space.
Algebraic consistency is ensured by the fact, that we are given an explicit representation.
The second approach of Brei\-tenlohner and Maison,
which is an axiomatic approach, treats these objects as abstract
symbols satisfying a certain set of (consistent) relations.
Within this approach it is important to verify the algebraic
consistency, since no explicit representation is given. 
In this paper we follow the lines of the constructive approach.\\
\\
We continue with an overview of existing regularization schemes within dimensional
regularization:
A scheme, which maintains the anticommuting property of $\gamma_5$, but gives up
the cyclicity of the trace was advocated in \cite{Kreimer}.\\
\\
Dimensional reduction \cite{DR} was formulated in an attempt to obtain a scheme
which preserves supersymmetry. It has been shown \cite{Bonneau} that it runs into
similar problems as the naive anticommuting scheme.
Dimensional reduction continues the momenta to $d < 4$ dimensions, but keeps spinors and vector fields in
four dimensions.
A vector field $A_\mu$ is split into a part
$A_i$ with $0 \le i \le (d-1)$, which transforms as a $d$-dimensional vector and
$A_\sigma$ with $d \le \sigma \le 3$, which transforms as a $(4-d)$-dimensional scalar.
Within dimensional reduction one has to distinguish between the $d$-dimensional
metric tensor $g_{\mu\nu}$ and the four-dimensional metric tensor, denoted by
$\tilde{g}_{\mu\nu}$. $g_{\mu\nu}$ acts as an orthogonal projection operator onto
the $d$-dimensional subspace:
\bq
\label{projection}
\tilde{g}_{\mu\rho} g^{\rho}_{\;\;\;\nu} & = & g_{\mu\nu}.
\eq
In dimensional reduction we have
\bq
\left\{ \gamma_\mu, \gamma_\nu \right\} & = & 2 \tilde{g}_{\mu\nu} \cdot 1,
\eq
since the Dirac algebra is in four dimensions. Since $g_{\mu\nu}$ is a projection onto the $d$-dimensional
space we obtain
\bq
g^{\mu\nu} \gamma_\mu \gamma_\nu & = & d \cdot 1.
\eq
If we consider now
\bq
\eps^{\mu\nu\rho\sigma} g^{\alpha\beta} \mbox{Tr}\;\left(\gamma_\alpha \gamma_\mu \gamma_\nu \gamma_\rho \gamma_\sigma \gamma_\beta \gamma_5 \right)
\eq
we arrive at the same contradiction as above.
Other inconsistencies of dimensional reduction have been reported by W. Siegel \cite{DR} and by
L.V. Avdeev and A.A. Vladimirov \cite{Avdeev}.\\
\\
The four-dimensional helicity scheme introduced by Z. Bern and D.A. Kosower \cite{FDH} is 
widely used in the calculaion of one-loop amplitudes
in massless QCD. It is only defined for parity-conserving
amplitudes, and does therefore not require the definition of a $\gamma_5$.
It is similar to dimensional reduction. 
In both schemes a massless vector particle (like the photon
or the gluon) has two helicity states. 
In dimensional reduction these two states are split
between a $d$-dimensional vector (which has $(d-2)$-states) and $(4-d)$ scalars which must be treated
separately.
In the FDH-scheme however, the two helicity states are carried by 
the $d$-dimensional vector particle.
The equivalence of the FDH scheme with dimensional reduction at the level of one-loop
has been shown by Z. Kunszt, A. Signer and Z. Tr\'ocs\'anyi \cite{split2}.
The FDH scheme leads to the same inconsistences as dimensional reduction, if it would be used for the
calculation of parity-violating loop amplitudes.\\
\\
One objective of this paper is the definition of a scheme, which in practical calculations
is as simple to use as the FDH-scheme, but which avoids the algebraic inconsistencies
associated with the later one.
We ensure algebraic consistency by explicit construction of our scheme.
Like in the 't Hooft--Veltman scheme we will encounter physical and non-physical
anomalies.
Non-physical anomalies have to be cancelled to ensure unitarity.
They arise when deformations due to the regularization scheme (which are of order
$\eps$) are combined with poles coming from the loop integration.
The poles of the loop integration are either ultraviolet or infrared in origin.
Non-physical anomalies from ultraviolet divergences lead to violations of Ward identities,
which must then be restored by hand at each order in perturbation theory.
C. Becchi, A. Rouet and R. Stora \cite{BRS} have shown that this can be done to all order
in perturbation theory if the theory is free of physical anomalies.
In addition there might be infrared divergences which are usually also regulated with 
the help of the same regularization scheme.
Again the question of unitarity arises. S. Catani, M.H. Seymour and Z. Tr\'ocs\'anyi \cite{split1} 
have shown how to recover unitarity and renormalization-scheme independence at the level of one-loop.
The necessarity of the restoration of the Ward identities is a drawback, which cannot
be avoided. However, since all algebra can be performed in four dimensions,
this allows the use of four-dimensional Schouten- or Fierz-identities in practical calculations.
There are situations where this advantage outweights the disadvantage coming from 
the necesarry restoration of the Ward idenities.\\
\\
In this paper we develop a general formalism, which allows us to define consistent regularization schemes.
We explicitly define two schemes, one similar
to the 't Hooft - Veltman scheme and the other one the proposed four-dimensional scheme. We do this in a Lorentz-invariant way.
In all intermediate steps we keep Lorentz-invariance, which motivates the word ``equivariant'' in the title.
(In the mathematical literature ``equivariant'' means roughly ``compatible with a group action''.
The group will be the Lorentz group $SO(1,3)$.)
However, we will show that, when defining spinors in the regularized theory, the Clifford algebra structure
cannot be preserved.
Now the two schemes, which we are going to define (one similar to the HV scheme and the other one the
four-dimensional scheme), correspond to two different deformations of the Clifford algebra structure. In essence, the HV-like
scheme deforms
\bq
\left\{ \gamma^\mu, \gamma_5 \right\},
\eq
whereas the four-dimensional scheme deforms
\bq
\left\{ \gamma^\mu, \gamma^\nu \right\}.
\eq
Consistency requires that the deformations are of order $O(\eps)$, so that tree-level amplitudes are not affected.
If these deformation effects are combined with poles arising from divergent integrals, they give rise to the well-known
violation of Ward identities, and anomalies.\\
\\
All regularization schemes discussed in the introduction involve continuing
space-time from four to $d$ dimensions. 
Likewise, 
the Lorentz group $G$, which acts on space-time, is always continued from $SO(1,3)$
to $SO(1,d-1)$. In this article we will, in contrast,
fix the Lorentz group to $SO(1,3)$, but instead of one vector representation and its vector space, 
consider a whole set of even-dimensional representations of $G$ and their vector spaces, on
which the definition of chirality is straightforward.\\
\\
To define integration over loop momenta on this set of spaces we take the following steps:
The set of all such even-dimensional representations forms abelian semi-groups with respect
to the direct sum and the tensor product. Using ideas inspired
by K-theory, we construct the associated abelian groups, which we combine into an object called
$Q(G)$. This $Q(G)$ is then an abelian group (``K-group'')
with respect to addition and multiplication.
With respect to the direct sum, the construction gives the well-known representative
ring of $G$, restricted to vector spaces of even dimensions.
Now, this K-group $Q(G)$ allows the definition of an homomorphism into the real numbers,
called the rank, such that the image of $Q(G)$ under that homomorphism
is dense in ${\Bbb R}$. The rank will play the r\^ole of the ``$4-2\eps$'' in conventional
dimensional regularization.
We then extend this construction to the complex case, leading to the complex equivalent of $Q(G)$,
$C(G)$, and a complex rank homomorphism.
Then, finally, integration on $C(G)$ of functions (of momenta) is defined such that it agrees with ordinary
integration on elements of $C(G)$ with positive integer rank.
\\
\\
This paper is organized as follows: The next section introduces the basic concepts
of K-theory, as far as they are needed in subsequent sections.
In section 3 we give the precise definition of integration in $d$ dimensions.
Section 4 deals with Dirac matrices in a space of even dimensions.
In section 5 and 6 we show how gauge groups or global symmetry groups are included into the regularization scheme.
As an example we do this explicitly for QCD in section 5. Section 6 indicates the extension for electroweak 
or supersymmetric theories.
Section 7 summarizes the definition of our regularization schemes.
In section 8 we apply our regularization schemes to various examples: The triangle
anomaly (AVV and VVA), the AAA-anomaly, the Ward identity for the non-singlet axial-vector current
and the Ward identity for the vector current are calculated.
Section 9 contains the conclusions. 
Appendix A contains an explanation of the notation used throughout this paper.
In the remaining appendices we have collected information, which we found too technical
to be included into the main text.

\section{Basic K-theory}

Given a Lie group $G$ we denote by
\bq
{\cal V} & = & \left\{ V_1, V_2, ..., V_i, ... \right\}
\eq
the set of all finite-dimensional representations of $G$.
The dimension of the vector space $V_i$ is denoted by
\bq
\mbox{dim} \; V_i & = & d_i
\eq
and is an integer number.
On ${\cal V}$  we have two operations, the direct sum and the tensor product, such that
\bq
V_i \oplus V_j \in {\cal V} & \mbox{and} & V_i \otimes V_j \in {\cal V}
\eq
are again elements of ${\cal V}$. It is easy to see that with respect to each of these operations
${\cal V}$ is an abelian semi-group.
The situation is similar to the natural numbers, which form also abelian semi-groups with respect to
addition and multiplication.
The dimensions of the resulting vector spaces are:
\bq
\mbox{dim} \left( V_i \oplus V_j \right) = d_i + d_j , & &
\mbox{dim} \left( V_i \otimes V_j \right) = d_i  d_j.
\eq
Later on we will restrict $\cal V$ to be the set of all finite-dimensional representations of even
dimensions. Since the addition or multiplication of two even numbers is again an even number, this restricted
set also forms abelian semi-groups with respect to the direct sum and the tensor product.
In the following we will construct spaces of negative or rational ``dimensions''. This is in complete
analogy of the constructions of the integer or rational numbers out of the natural numbers.\\
\\
The mathematical framework, which associates to each abelian semi-group an abelian group, is the domain
of K-theory. 
We first review briefly how this construction is carried out \cite{Brodzki}. 
Let $A$ be an abelian semi-group; we assume for simplicity
that it contains a zero element. The Grothendieck group $K(A)$ of $A$
is an abelian group that has the following universal property: There is a 
canonical semi-group homomorphism $\phi_A : A \rightarrow K(A)$ such that
for any group $G$ and semi-group homomorphism $\psi : A \rightarrow G$, there
is a unique homomorphism $\gamma : K(A) \rightarrow G$ such that $\psi =\gamma \phi_A$. This means that the following diagram is commutative:
\begin{center}
\begin{picture}(80,80)(0,0)
\Text(10,60)[cc]{$A$}
\Text(70,60)[cc]{$K(A)$}
\Text(40,20)[cc]{$G$}
\LongArrow(20,60)(55,60)
\LongArrow(15,55)(35,25)
\LongArrow(65,55)(45,25)
\Text(40,65)[cb]{$\phi_A$}
\Text(15,40)[cc]{$\psi$}
\Text(65,40)[cc]{$\gamma$}
\end{picture}
\end{center}
To prove the existence of $K(A)$ we provide three constructions of $K(A)$:
\begin{enumerate}
\item Let $F(A)$ be the free abelian group generated by the elements of $A$, and let $E(A)$
be the subgroup of $F(A)$ generated by the elements of the form $a+b-(a \oplus b)$, where
$\oplus$ denotes addition in $A$. We define
\bq
K(A) & = & F(A) /E(A)
\eq
with $\phi_A : A \rightarrow K(A)$ being the composition of the inclusion $A \rightarrow F(A)$ with the
canonical surjection $F(A) \rightarrow K(A)$.
\item Let $\Delta : A \rightarrow A \times A$ be the diagonal homomorphism of semi-groups, 
e.g. an element $a \in A$ is mapped to $(a,a) \in A \times A$,
and let
$K(A)$ be the set of cosets of $\Delta(A)$ in $A \times A$. A priori it is only a quotient
semi-group, but it is not difficult to check that the interchange of factors in $A \times A$ induces
an inverse in $K(A)$ so that $K(A)$ is in fact a group. We then define $\phi_A : A \rightarrow K(A)$ to be
 the composition of the map $a \rightarrow (a,0)$ with the natural projection $A \times A \rightarrow K(A) $.
\item Consider the following equivalence relation on the set-theoretical product $A \times A$. We put
$(a,b) \sim (a',b')$ when there exists a $p \in A$ such that
\bq
a + b' + p & = & a' + b +p.
\eq
Then by definition $K(A) = A \times A / \sim$. Elements of $K(A)$ will be denoted $\left[(a,b)\right]$.  
The additional $p$ is needed in order to ensure transitivity of the equivalence relation.
If $(a_1,b_1) \sim (a_2,b_2)$ and $(a_2,b_2) \sim (a_3,b_3)$ we have a $c_1$ and a $c_2$ such that
\bq
a_1 + b_2 + c_1 & = & a_2 + b_1 + c_1, \\
a_2 + b_3 + c_2 & = & a_3 + b_2 + c_2.
\eq
It follows that
\bq
a_1 + b_3 + c_3 & = & a_3 + b_1 + c_3
\eq
with $c_3 = a_2 + b_2 + c_1 + c_2$.
\end{enumerate}
For example the integer numbers are constructed as follows:
We consider pairs $(a,b)$ with $a,b \in {\Bbb N}_0$ and an addition
defined by components
\bq
(a_1,b_1) + (a_2,b_2) & = & (a_1+a_2,b_1+b_2).
\eq
Furthermore we have the additional relation that two elements $(a_1,b_1)$
and $(a_2,b_2)$ are equivalent if
\bq a_1 + b_2 + n = a_2 + b_1 + n
\eq
for some $n \in {\Bbb N}_0$. 
Equivalence classes are then denoted by $[(a,b)]$.
The inverse of $[(a,b)]$ is $[(b,a)]$.\\
\\
In order to obtain an inverse to the direct sum operation on our set
of vector spaces ${\cal V}$ we proceed similar: We consider pairs
$(V_i,V_j)$ with $V_i, V_j \in {\cal V}$, define the addition by components
\bq
\left(V_{i_1},V_{j_1}\right) + \left(V_{i_2},V_{j_2}\right) & = &
\left( V_{i_1} \oplus V_{i_2}, V_{j_1} \oplus V_{j_2} \right)
\eq
and call two elements $(V_{i_1},V_{j_1})$ and $(V_{i_2},V_{j_2})$ 
equivalent if there is a $V_k$ such that
\bq
V_{i_1} \oplus V_{j_2} \oplus V_k & \mbox{and} &
V_{i_2} \oplus V_{j_1} \oplus V_k
\eq
are isomorph. We denote the equivalence classes by
$[(V_i,V_j)]$. The group $K_G({\cal V})$ is usually called the representative
ring of $G$ and denoted by
\bq
R(G) & = & K_G({\cal V}).
\eq
We define the rank 
of an element $[(V_i,V_j)]$ by
\bq
\mbox{rank}\; [(V_i,V_j)] & = & \mbox{dim}\; V_i - \mbox{dim}\; V_j
\eq
which is an integer number.
It is easy to check that the rank does not depend on the representative.\\
\\
We note that if we choose a specific pair $(V_i,V_j)$ to represent the equivalence class $[(V_i,V_j)]$,
this pair is a vector space of dimension $d_i + d_j$. 
However, different representatives for the same
equivalence class may have different dimensions, when viewed as a vector space.
The quantity which is independent of the chosen representative is the rank defined above.
The radial variable in the vectorspace $(V_i,V_j)$
is given by
\bq
k^2 & = & k_i^2 + k_j^2,
\eq
where $k_i$ and $k_j$ are the radial variables of $V_i$ and $V_j$, respectively.\\
\\
In order to construct rational numbers and spaces of rational rank we apply the K-functor again,
this time with respect to multiplication and the tensor product: The rational numbers are
constructed by considering pairs $(z_a,z_b)$, where $z_a$ and $z_b$ are now elements of ${\Bbb Z}\setminus\{0\}$.
We may think of $z_a$ as the numerator and of $z_b$ as the denominator of a rational number $z_a/z_b$.
The multiplication is defined by components:
\bq
\left(z_{a_1},z_{b_1}\right) \cdot \left(z_{a_2},z_{b_2} \right) & = &
\left( z_{a_1} z_{a_2}, z_{b_1} z_{b_2} \right)
\eq
and two elements $(z_{a_1},z_{b_1})$ and $(z_{a_2},z_{b_2} )$
are called equivalent if
\bq
z_{a_1} z_{b_2} z_c & = & z_{a_2} z_{b_1} z_c
\eq
for some $z_c \in {\Bbb Z}\setminus \{0\}$.\\
\\
We construct spaces of rational rank as follows: We consider pairs
$( [(V_{i_1},V_{j_1})], [(V_{i_2},V_{j_2})])$, where $[(V_{i_k},V_{j_K})] \in R(G)$, define the multiplication
by components:
\bq
\lefteqn{\left( \left[\left(V_{i_1},V_{j_1}\right)\right], \left[ \left(V_{i_2},V_{j_2} \right) \right] \right)
\cdot
\left( \left[ \left(V_{i_3},V_{j_3}\right) \right] , \left[ \left(V_{i_4},V_{j_4} \right) \right] \right)
= } & & \nonumber \\
& = & \left( \left[ \left(V_{i_1},V_{j_1} \right) \right] \otimes \left[ \left(V_{i_3},V_{j_3}\right)\right], 
\left[ \left(V_{i_2},V_{j_2} \right) \right] \otimes \left[\left(V_{i_4},V_{j_4}\right)\right]\right) 
\eq
where
\bq
\left[\left(V_{i_1},V_{j_1}\right)\right] \otimes \left[\left(V_{i_3},V_{j_3}\right)\right] & = &
\left[\left( \left( V_{i_1} \otimes V_{i_3} \right) \oplus \left( V_{j_1} \otimes V_{j_3} \right) ,
\left( V_{i_1} \otimes V_{j_3} \right) \oplus \left( V_{i_3} \otimes V_{j_1} \right) \right)\right].
\eq
Two elements, $( [(V_{i_1},V_{j_1})], [(V_{i_2},V_{j_2})])$ and $( [(V_{i_3},V_{j_3})], [(V_{i_4},V_{j_4})])$,
are equivalent if there is an element $[(V_{i_5},V_{j_5})] \in R(G)$ such that
\bq
\left[\left(V_{i_1},V_{j_1}\right)\right] \otimes \left[\left(V_{i_3},V_{j_3}\right)\right]
\otimes \left[\left(V_{i_5},V_{j_5}\right)\right]
\eq
and
\bq
\left[\left(V_{i_2},V_{j_2}\right)\right] \otimes \left[\left(V_{i_4},V_{j_4}\right)\right]
\otimes \left[\left(V_{i_5},V_{j_5}\right)\right]
\eq
are isomorph. We will call the resulting K-group $K_G(R(G))$ the representative field of $G$ and denote
it by
\bq
Q(G) & = & K_G\left( R(G) \right).
\eq
Elements of $Q(G)$ are denoted by
\bq
\left[ \left( \left[\left(V_{i_1},V_{j_1}\right)\right], \left[ \left(V_{i_2},V_{j_2} \right) \right] \right) \right].
\eq
This notation is rather cumbersome and we will introduce the more suggestive notation
\bq
\frac{V_{i_1} - V_{j_1}}{V_{i_2}-V_{j_2}}.
\eq
The rank of an object in $Q(G)$ is given by
\bq
\mbox{rank}\;\left( \frac{V_{i_1} - V_{j_1}}{V_{i_2}-V_{j_2}}\right) & = & \frac{d_{i_1} - d_{j_1}}{d_{i_2} - d_{j_2}}.
\eq
We note that a particular representative of an element in $Q(G)$ is a vector space of dimension
$(d_{i_1}+d_{j_1})(d_{i_2}+d_{j_2})$. 
(As a vector space it is just the tensor product of $(V_{i_1} \oplus V_{j_1})$ with
$(V_{i_2} \oplus V_{j_2})$.)
As already mentioned above, different representatives for the same
equivalence class may have different dimensions, when viewed as a vector space.
The quantity which is independent of the chosen representative is the rank defined above.\\
\\
We may think of $V_i/V_j$ as the the equivalence class of $V_i \otimes V_j^\ast$, where
$V_j^\ast$ is the dual space of linear forms on $V_j$, together with the convention
that
\bq
V_j^\ast \otimes V_j \cong {\Bbb C},
\eq
e.g. $V_j^\ast$ acts on $V_j$ whenever there is a $V_j$ in the product.
Then 
\bq
\left( V_i \otimes V_j^\ast \right) \otimes V_j &= & V_i.
\eq
A word of warning on the notation: The notation $V_i/V_j$ is a short-hand notation for
the equivalence
class $[([(V_i,\emptyset)],[(V_j,\emptyset)])]$.
It does not denote a coset space.
Since we never use cosets in this article, the notation should be clear.\\
\\
Spaces of complex rank are defined as follows: We consider pairs
\bq
\left( W_1, W_2 \right)
\eq
with $W_1,W_2 \in Q(G)$, define the addition by components and the multiplication
by
\bq
\left( W_1, W_2 \right) \cdot \left( W_3, W_4 \right) & = & \left(
W_1 \cdot W_3 - W_2 \cdot W_4, W_1 \cdot W_4 + W_2 \cdot W_3 \right).
\eq
We will denote the corresponding field by $C(G)$.
The rank of $(W_1,W_2) \in C(G)$ is defined as
\bq
\mbox{rank}\;\left( W_1, W_2 \right) & = & \mbox{rank}\; W_1 + i \;\mbox{rank}\; W_2.
\eq
It is easily checked that $C(G)$ is indeed a field and that the rank function is a homomorphism
\bq
\mbox{rank} & : & C(G) \rightarrow {\Bbb C}.
\eq
We may therefore use the following short-hand notation for objects of $C(G)$:
\bq
\left( W_1, W_2 \right) = W_1 + i W_2
\eq
An element of $C(G)$ can be represented by an octuplet of vector spaces,
where each vector space is a representation of $G$.
We write for an object of $X \in C(G)$ symbolically
\bq
X & = & \frac{V_1+V_2}{V_3+V_4} + i \frac{V_5+V_6}{V_7+V_8}
\eq
with $V_i \in {\cal V}$.

\section{Definition of the integration}

We proceed to define an integration over elements $X \in C(G)$. 
A representative for $X$ is a vectorspace, and we assume that its dimension is $(2m)$.
In general the dimension of the representative as a vector space does not equal the rank of $X$.\\
\\
We restrict the set of functions, which we allow to be integrated: We shall only consider
functions $f$ , which depend on the coordinates through $k^2$ and  $(2 k q)$:
\bq
f & = & f\left( k^2, 2 k q \right),
\eq
where $k^2$ denotes the square of the radial variable in $X$, and 
$(2 k q)$ denotes symbolically a collection of scalar products of the loop
momentum $k$ with some external momenta $q$. \\
\\
For a given function $f$ we define the integration over $X \in C(G)$ by
\bq
I & : & C(G) \rightarrow {\Bbb C}, \nonumber \\
& & X \rightarrow \int \left( d^{\;\mbox{rank}\;X}k\right) \;\;f.
\eq
This definition has two parts: First of all, given a vectorspace of dimension $(2m)$ and a function $f$, defined
on that vectorspace $X$ and satisfying certain restrictions, we have to define the $d$-dimensional integral
of this function $f$ over the space $X$. Note that we do not require that $d$ is equal to the dimension of $X$, 
which is $(2m)$. 
The appropriate definition is well known within the context of dimensional regularization \cite{Wilson,CDR}.
Having done this we may now choose $X$ to be an element of $C(G)$ and set $d$ equal to the rank
of $X$. The subtle point is, that we have to show that the definition is independent of the 
representative for $X$.\\
\\
We start with the first point: 
Given a vectorspace $X$ of dimension $(2m)$ and a function $f$ which only
depends on $k^2$ and $(2 k q)$,
we first further assume that $f \rightarrow 0$ rapidly enough as $k^2 \rightarrow \infty$ and that $f$ is
analytic
for $k^2 =0$.
Then we define the integral over $X$ for $\mbox{Re}\;d > 2m$ by
\bq
\label{definition1}
\int d^dk f(k^2, 2kq) & = & \frac{\pi^{d/2-m}}{\Gamma\left(\frac{d}{2}-m\right)} \int d^{2m}k \int dk_\perp^2
\left( k_\perp^2 \right)^{d/2-m-1} f\left( k^2+k_\perp^2, 2 k q \right).
\eq
For all other values of $d$ the integral is then defined by analytic continuation.
If $f$ goes not rapidly enough to zero as $k^2 \rightarrow \infty$ or if $f$ has a singularity
at $k^2=0$, we refer for details to the book by Collins \cite{CDR}.\\
\\
The definition of the integration satisfies \cite{Wilson,CDR}:
\begin{enumerate}
\item Linearity: For two functions $f_1$ and $f_2$ and two constants $a$ and $b$ we have
\bq
\int d^dk \left( a f_1 + b f_2 \right) & = & a \int d^dk f_1 + b \int d^dk f_2.
\eq
\item Translation invariance:
\bq
\int d^dk f(k+q) & = & \int d^dk f(k)
\eq
\item Scaling law:
\bq
\int d^dk f(\lambda k) & = & \lambda^{-d} \int d^dk f(k)
\eq
\item Normalization: 
\bq
\int d^dk e^{-k^2} & = & \pi^{d/2}
\eq
\end{enumerate}
Having established translation invariance, we may use Feynman parametrization and a shift in the loop
momentum, such that the integrand depends on the radial variable $k^2$ only.
In that case the definition of $d$-dimensional integration reduces to
\bq
\label{definition2}
\int d^dk f(k^2) & = & \frac{\pi^{d/2}}{\Gamma\left(\frac{d}{2}\right)} \int dk^2 \left(k^2\right)^{d/2-1}
f(k^2).
\eq
We now consider how $d$-dimensional integration behaves under addition and multiplication of elements of $C(G)$.
We first consider $(d_1+d_2)$-dimensional integration over $X_1 + X_2$, where $X_1,X_2 \in C(G)$.
We want to show:
\bq
\label{proof_addition}
\int\limits_{X_1+X_2} d^{d_1+d_2}k f(k^2) & = & \int\limits_{X_1} d^{d_1}k_1 \int\limits_{X_2} d^{d_2}k_2 f(k_1^2+k_2^2) \nonumber \\
& = & \int\limits_{X_2} d^{d_2}k_2 \int\limits_{X_1} d^{d_1}k_1 f(k_1^2+k_2^2),
\eq
where we have written $k^2 = k_1^2 + k_2^2$.
In order to prove this, we apply the definition eq. (\ref{definition1}):
\bq
\label{calc1}
\int\limits_{X_1} d^{d_1}k_1 \int\limits_{X_2} d^{d_2}k_2 f(k_1^2+k_2^2) & = & 
\frac{\pi^{d_1/2+d_2/2-m_1-m_2}}{\Gamma\left(\frac{d_1}{2}-m_1\right)\Gamma\left(\frac{d_2}{2}-m_2\right)}
\int d^{2m_1+2m_2}k \int dk_{1\perp}^2 \int dk_{2\perp}^2 \nonumber \\
& & \cdot \left(k_{1\perp}^2\right)^{d_1/2-m_1-1}
\left(k_{2\perp}^2\right)^{d_2/2-m_2-1}
f \left(k^2 + k_{1\perp}^2 + k_{2\perp}^2 \right) 
\eq
We then perform a change of variables according to
\bq
t = \frac{k_{1\perp}^2}{k_{1\perp}^2 + k_{2\perp}^2}, & & k_\perp^2 = k_{1\perp}^2 + k_{2\perp}^2.
\eq
The Jacobian gives a factor
\bq
\left| \frac{\partial(k_{1\perp}^2,k_{2\perp}^2)}{\partial(t,k_{\perp}^2)}
\right| &=  & k_\perp^2 .
\eq
The integral over $t$ yields
\bq
\int\limits_0^1 dt \; t^{d_1/2-m_1-1} \left( 1 - t \right)^{d_2/2-m_2-1} & = & 
\frac{\Gamma\left( \frac{d_1}{2}-m_1 \right) \Gamma\left( \frac{d_2}{2}-m_2 \right)}
{\Gamma\left( \frac{d_1}{2}+\frac{d_2}{2}-m_1-m_2 \right)}
\eq
and therefore eq. (\ref{calc1}) equals
\bq
\int\limits_{X_1+X_2} d^{d_1+d_2}k f(k^2),
\eq
In a similar way we can also obtain the following stronger result: If $f$ depends on $k_1^2$ and on $k_2^2$ separately, but
not on $2 k_1 k_2$, then the order of integration does not matter:
\bq
\int d^{d_1}k_1 \int d^{d_2}k_2 f(k_1^2,k_2^2) & = & \int d^{d_2}k_2 \int d^{d_1}k_1 f(k_1^2,k_2^2)
\eq
In particular if $k_1$ and $k_2$ belong to orthogonal vectorspaces, $f$ will not depend on
$2 k_1 k_2$. This fact will be useful for the calculation of two loops and beyond.\\
\\
Next we consider $(d_1 d_2)$-dimensional integration over $X_1 \cdot X_2$. We assume that $X_1$ and $X_2$
are given by vector spaces of dimensions $2m_1$ and $2m_2$, respectively. $X_1 \cdot X_2$ is then a vector space 
of dimension $4 m_1 m_2$ and the integration gives according to eq. (\ref{definition1}) and eq. (\ref{definition2}):
\bq
\label{tensor}
\int\limits_{X_1 \cdot X_2} d^{d_1 d_2} k f(k^2) & = & 
\frac{\pi^{d_1d_2/2-2m_1m_2}}{\Gamma\left( \frac{d_1d_2}{2}-2m_1m_2 \right)}
\int d^{4m_1 m_2}k \int dk_\perp^2 \left( k_\perp^2 \right)^{d_1d_2/2-2m_1m_2-1} f \left( k^2 + k_\perp^2 \right) \nonumber \\
& = & \frac{\pi^{d_1d_2/2}}{\Gamma\left(\frac{d_1d_2}{2}\right)} \int dk^2 \left( k^2 \right)^{d_1d_2/2-1} f(k^2)
\eq
For the addition we have shown in eq. (\ref{proof_addition}) that integration over $X_1+X_2$ gives the same result as 
first performing the integration over $X_2$ and then performing the integration over $X_1$.
A similar statement can be made for the multiplication.
The detailed calculation, where the integration is first performed over
``$ (X_1 \cdot X_2) \; \mbox{mod}\;X_1$'' and then over $X_1$, is given in appendix D and gives indeed the same result
as eq. (\ref{tensor}).\\
\\
We now come to the second point. We have to show that the definition of integration
\bq
\int \left( d ^{\mbox{rank}\;X} k \right) f
\eq
is well defined, e.g. does not depend on the vectorspace, which we choose as representative
for $X$.
Clearly $(\mbox{rank}\;X)$ is by construction independent of the representative.
However there is also an implicit dependence on the representative through the function $f$, which
is given as a function defined on the vectorspace, which we have chosen as a representative.
We have to show that the integral is independent of that.
We restrict ourselves to functions, which depend on the radial variable only.
We have to prove:
\bq
\int\limits_{(X_1+X_3)-(X_2+X_3)} dk f(k^2) & = & \int\limits_{X_1-X_2} dk f(k^2), \\
\int\limits_{(X_1 \cdot X_3)/(X_2 \cdot X_3)} dk f(k^2) & = & \int\limits_{X_1/X_2} dk f(k^2). \label{quotient2}
\eq
We start with addition:
\bq
\lefteqn{\int\limits_{(X_1+X_3)-(X_2+X_3)} dk f(k^2) = \frac{\pi^{m_1-m_2}}{\Gamma(m_1+m_3)\Gamma(-m_2-m_3)} } &  & \nonumber \\
& & \cdot
\int\limits_0^\infty dk_{(1+3)}^2 \int\limits_0^\infty dk_{(2+3)}^2 \left( k_{(1+3)}^2\right)^{m_1+m_3-1} \left( k_{(2+3)}^2\right)^{-m_2-m_3-1}
f(k_{(1+3)}^2 + k_{(2+3)}^2) \nonumber \\
& = & \frac{\pi^{m_1-m_2}}{\Gamma(m_1+m_3)\Gamma(-m_2-m_3)} 
\int\limits_0^1 dt \int\limits_0^\infty dk^2 \left(k^2 \right)^{m_1-m_2-1} t^{m_1+m_3-1} 
\left( 1 -t \right)^{-m_2-m_3-1} f(k^2) \nonumber \\
& = & \frac{\pi^{m_1-m_2}}{\Gamma(m_1-m_2)} \int\limits_0^\infty dk^2 \left(k^2 \right)^{m_1-m_2-1}
f(k^2) \nonumber \\
& = & \int\limits_{X_1-X_2} dk f(k^2) 
\eq
Here we used the notation that $k_{(1+3)}^2$ is the square of the radial variable of the vector space
$X_1 + X_3$, and similar for $k_{(2+3)}^2$.\\
\\
For the multiplication we also have to check that
integration over $X_1/X_2$ and
integration over $(X_1 \cdot X_3)/(X_2 \cdot X_3)$ yield the same result. 
Now
\bq
\int\limits_{X_1/X_2} dk f(k^2) & = & \frac{\pi^{m_1/2m_2}}{\Gamma\left(\frac{m_1}{2m_2}\right)}
\int dk^2 \left( k^2 \right)^{m_1/2m_2 -1 } f(k^2).
\eq
On the other hand we obtain
\bq
\int\limits_{(X_1 \cdot X_3)/(X_2 \cdot X_3)} dk f(k^2) & = & 
\frac{\pi^{\frac{m_1 m_2}{2 m_2 m_3}}}{\Gamma\left(\frac{m_1 m_2}{2 m_2 m_3}\right)} \int dk^2 \left(k^2
\right)^{\frac{m_1 m_2}{2 m_2 m_3}-1} f(k^2) \nonumber \\
& = & \frac{\pi^{m_1/2m_2}}{\Gamma\left(\frac{m_1}{2m_2}\right)}
\int dk^2 \left( k^2 \right)^{m_1/2m_2 -1 } f(k^2).
\eq
Therefore the result of the integration does not depend on the representative and
the integration is well defined.\\
\\
In summary the integration over an element of $X \in C(G)$ of rank $r$ is defined by
\bq
\int\limits_{X} dk f(k^2) & = & \frac{\pi^{r/2}}{\Gamma(r/2)}
\int\limits_0^\infty dk^2 \left( k^2 \right)^{r/2-1} f(k^2).
\eq
Since the image of $C(G)$ under the rank homomorphism is a dense subset 
\footnote{This is quite a general statement and does not depend on the specific structure of $G$. For any $G$ 
with a finite-dimensional representation $V$ of $G$, we can first construct $X=V/V$, which is of rank 1. We then can construct
an object $(pX)/(qX)$ of rank $p/q$, where $p,q \in {\Bbb Z}$ and $pX=X+X+...+X$ ($p$ times).}
of ${\Bbb C}$ (this derives from the
fact that
$\Bbb Q$ is dense in $\Bbb R$), we can find for every $d \in {\Bbb C}$ and every $\delta > 0$
a $X \in C(G)$ such that
\bq
\left| d - \;\mbox{rank}(X) \right| < \delta.
\eq
Furthermore, since $\int f$ is analytic in some domain $D$ of the complex plane we can find for every $\delta_1 >0$ a $\delta_2 >0$
such that
\bq
\left| I(X_i) - I(X_j) \right| < \delta_1 
\eq
for all $X_i$, $X_j \in C(G)$ with
\bq
\left| \mbox{rank}(X_i) - \mbox{rank}(X_j) \right| < \delta_2
\eq
and
\bq
\mbox{rank}(X_i) \in D & \mbox{and} & \mbox{rank}(X_j) \in D.
\eq
In practice we will regulate one-loop integrals by integration over
\bq
X_0 + X_1.
\eq
$X_0$ has rank 4 and is represented by the physical Minkowski space. $X_1$ is of rank $(-2\eps)$ and
serves to regulate the loop.
Multi-loop amplitudes are calculated by integration over 
\bq
X_{0,l-1} + X_l,
\eq
where $X_l$ regulates the $l$-th loop and $X_{0,l-1}$ contains the physical space $X_0$ as well as all
the $X_{j}$ (with $1 \le j \le l-1$) needed to regulate the remaining $(l-1)$-loops:
\bq
X_{0,l-1} & = & X_0 + X_1 + ... + X_{l-1}
\eq
We note that a vector
in $X_l$ is always orthogonal to all vectors of $X_j$ with $j<l$.\\
\\
We now have to define Dirac spinors over $X_{0,l}$. Since every $X_j$ is constructed out of
vectorspaces of even dimension, each representative of $X_j$ has even dimensions as well.
In the next section we deal therefore with Dirac matrices in spaces of even dimensions. 

\section{Dirac matrices in even dimensions}

We now construct a chiral representation of the Dirac matrices in a space of even dimensions $d = 2m$.
In even dimensions $d = 2 m$, the standard representation
of the $\gamma^\mu$'s has dimension $2^m$.
The Dirac matrices  satisfy the 
anticommutation relation
\bq
\label{anticom}
\left\{ \gamma^\mu_{(m)} , \gamma^\nu_{(m)} \right\} & = & 2 g^{\mu\nu}_{(m)} 1_{(m)}
\eq
and the hermitian requirement
\bq
\label{hermitian}
\left( \gamma^0_{(m)} \right)^\dagger & = & \gamma^0_{(m)}, \nonumber \\
\left( \gamma^i_{(m)} \right)^\dagger & = & - \gamma^i_{(m)}, \;\;\; 1 \le i \le 2m-1.
\eq
The index $m$ labels (one half of) the dimension of the space.
In even dimension we can further define a $\gamma_5$, which we shall denote by $\chi_{(m)}$. It satisfies
\bq
\label{gamma5}
\chi_{(m)}^2 & = & 1, \nonumber \\
\chi_{(m)}^\dagger & = & \chi_{(m)}, \nonumber \\
\left\{ \gamma^\mu_{(m)}, \chi_{(m)} \right\} & = & 0.
\eq
For $m=1$, e.g. $d=2$ we define
\bq
\gamma^0_{(1)} = - \sigma_y = \left( \begin{array}{cc}
0 & i \\ 
-i & 0 \\ \end{array}
\right),
& &
\gamma^1_{(1)} = i \sigma_x = \left( \begin{array}{cc}
0 & i \\ 
i & 0 \\ \end{array}
\right)
\eq
and define $\chi_{(m)}$ by
\bq
\chi_{(m)} & = & - \left(-i\right)^{m-1} \gamma^0_{(m)} ... \gamma^{2m-1}_{(m)},
\eq
e.g. $\chi_{(1)}$ is given by
\bq\chi_{(1)} = \sigma_z = \left(\begin{array}{cc}
1 & 0 \\
0 & -1 \\
\end{array}\right).
\eq
Given a representation for a given $m$, we construct a representation for $(m+1)$ as follows:
\bq
\gamma^0_{(m+1)} & = & \left( \begin{array}{cc}
                             0 & 1_{(m)} \\ 
                             1_{(m)} & 0 \\
                       \end{array} \right), \nonumber \\
\gamma^j_{(m+1)} & = & \left( \begin{array}{cc}
                             0 & i \gamma^j_{(m)} \\ 
                             -i \gamma^j_{(m)} & 0 \\
                       \end{array} \right) ,\;\;\; 1 \le j \le 2m-1, \nonumber \\
\gamma^{2m}_{(m+1)} & = & \left( \begin{array}{cc}
                             0 & \gamma^0_{(m)} \\ 
                             -\gamma^0_{(m)} & 0 \\
                       \end{array} \right), \nonumber \\
\gamma^{2m+1}_{(m+1)} & = & \left( \begin{array}{cc}
                             0 & -\chi_{(m)} \\ 
                             \chi_{(m)} & 0 \\
                       \end{array} \right). 
\eq
It is easily checked by induction that the anticommutation relation eq.(\ref{anticom}), the hermitian
requirement eq.(\ref{hermitian}) and the properties of $\chi_{(m)}$ in eq.(\ref{gamma5}) are fullfilled.
Furthermore we can show that in this representation we have for all $m$
\bq
\chi_{(m)} & = & \left( \begin{array}{cc}
                             1_{(m)} & 0 \\ 
                             0 & -1_{(m)} \\
                       \end{array} \right).
\eq
The Dirac matrices $\gamma^\mu_{(m)}$ act on a complex vector space of dimension $2^m$.
It is obvious that
\bq
\label{projectors}
P_{(m)}^+ = \frac{1}{2} \left( 1 + \chi_{(m)} \right) & \mbox{and} &
P_{(m)}^- = \frac{1}{2} \left( 1 - \chi_{(m)} \right)
\eq
are projection operators on the first $2^{m-1}$ components and on the last $2^{m-1}$ components, respectively.
We may therefore write any spinor as
\bq
\psi_{(m)} & = & \left( \begin{array}{c}
\chi_{(m)\;A} \\ \xi^{\dot{B}}_{(m)} \\
\end{array}
\right),
\eq
where the indices $A$ and $\dot{B}$ run from 1 to $2^{m-1}$.\\
\\
The definitions are such that for $m=2$ we have the well-known Weyl representation for the Dirac
matrices:
\bq
\gamma^{\mu}_{(2)} = \left(\begin{array}{cc}
 0 & \sigma^{\mu}_{(2)} \\
 \bar{\sigma}^{\mu}_{(2)} & 0 \\
\end{array} \right),
\eq
where $\sigma_{(2)\;A \dot{B}}^{\mu} = \left( 1 , - \sigma_{(2)}^i \right)$
and $\bar{\sigma}^{\mu \dot{A} B}_{(2)} = \left( 1 ,  \sigma_{(2)}^i \right)$.
The $\sigma_{(2)}^i$ are the Pauli matrices. \\
\\
We may cast our results for all $m$ in the same form and define $2^{m-1} \times 2^{m-1}$ matrices $\sigma^\mu_{(m)\;A \dot{B}}$
and $\bar{\sigma}^{\mu\;\dot{A}B}_{(m)}$ such that the relation above holds for all $m \in {\Bbb N}$.
Explicitly we obtain
\bq
\sigma_{(m+1)}^j & = & -i \gamma_{(m)}^j, \;\;\;1 \le j \le 2m-1, \nonumber \\
\sigma_{(m+1)}^{2m} & = & - \gamma_{(m)}^0, \nonumber \\
\sigma_{(m+1)}^{2m+1} & = & \chi_{(m)}.
\eq
Then
\bq
\sigma^\mu_{(m)\;A \dot{B}} = \left( 1, - \sigma_{(m)}^j \right), & & 
\bar{\sigma}^{\mu\;\dot{A}B}_{(m)} = \left( 1, \sigma_{(m)}^j \right)
\eq
and
\bq
\gamma^{\mu}_{(m)} = \left(\begin{array}{cc}
 0 & \sigma^{\mu}_{(m)} \\
 \bar{\sigma}^{\mu}_{(m)} & 0 \\
\end{array} \right).
\eq
In the following we use the notation
\bq
p\!\!\!/_{(m)} & = & \sum\limits_{\mu=0}^{2m-1} p^{(m)}_\mu \gamma^\mu_{(m)}, \\
(p q)_{(m)} & = & 2 \sum\limits_{\mu=0}^{2m-1} p^{(m)}_\mu q_{(m)}^\mu .
\eq
We state a few properties of traces of Dirac matrices in $(2m)$-dimensions:
\begin{enumerate}
\item The trace of an odd number of Dirac matrices equals zero:
\bq
\mbox{Tr} \; a\!\!\!/^{(m)}_1 ... a\!\!\!/^{(m)}_{2n+1} & = & 0
\eq
This can be proven by inserting $\chi_{(m)}^2$, using the anticommutation relation of $\chi_{(m)}$ and the
cyclic property of the trace.
\item The trace of an even number of Dirac matrices gives
\bq
\label{tr_even}
\lefteqn{
\mbox{Tr} \; a\!\!\!/^{(m)}_1 ... a\!\!\!/^{(m)}_{2n} =  
\frac{1}{2} \left( a_1 a_2 \right)_{(m)} \mbox{Tr}\; a\!\!\!/_3^{(m)} ... a\!\!\!/^{(m)}_{2n} } & & \nonumber \\
& & - \frac{1}{2} \left( a_1 a_3 \right)_{(m)} \mbox{Tr}\; a\!\!\!/_2^{(m)} a\!\!\!/_4^{(m)} ... a\!\!\!/^{(m)}_{2n}
+ ... + \frac{1}{2} \left( a_1 a_n \right)_{(m)} \mbox{Tr}\; a\!\!\!/_2^{(m)} ... a\!\!\!/^{(m)}_{2n-1}.
\eq
This is proven using the anticommutation relation $\left\{\gamma^\mu_{(m)}, \gamma^\nu_{(m)} \right\} = 2 g^{\mu\nu}_{(m)} 1_{(m)}$ and the cyclic property of the trace.
\item For $n<m$ we have
\bq
\mbox{Tr} \; a\!\!\!/^{(m)}_1 ... a\!\!\!/^{(m)}_{2n} \chi^{(m)} & = & 0.
\eq
This can be proven using eq. (\ref{tr_even}) and the fact that $\chi^{(m)}$ is the product of $(2m)$ Dirac
matrices. For $2n < 2m$ there are not enough indices in the string $a\!\!\!/^{(m)}_1 ... a\!\!\!/^{(m)}_{2n}$
to give a non-zero value.
\item For $n=m$ we find
\bq
\mbox{Tr}\; a\!\!\!/^{(m)}_1 ... a\!\!\!/^{(m)}_{2m} \chi^{(m)} & = & (i)^{m-1} \eps_{\mu_1 ... \mu_{2m}}
a_1^{\mu_1} ... a_{2m}^{\mu_{2m}} \mbox{Tr} \; 1^{(m)},
\eq
where $\eps_{\mu_1 ... \mu_{2m}}$ is the Levi-Civita symbol in $(2m)$-dimensions.
\end{enumerate}
We finally consider the case of Dirac matrices in $2(m_1+m_2)$-dimensions. With the definition
\bq
\chi_{(m_1,m_1+m_2)} & = & - (-i)^{m_1-1} \gamma^0_{(m_1+m_2)} ... \gamma^{2m_1-1}_{(m_1+m_2)}
\eq
the matrices $1_{(m_1+m_2)}$, $\chi_{(m_1,m_1+m_2)}$ and $\gamma^\mu_{(m_1+m_2)}$ with $0 \le \mu \le 2m_1-1$ clearly
form a representation of an $2m_1$-dimensional Dirac algebra.
The anticommutation relation of
$\gamma^\mu_{(m_1+m_2)}$ with $\chi_{(m_1,m_1+m_2)}$ for $2m_1 \le \mu \le 2m_1+2m_2-1$ reads:
\bq
\left\{ \gamma^\mu_{(m_1+m_2)}, \chi_{(m_1,m_1+m_2)} \right\} & = & 
2 \gamma^\mu_{(m_1+m_2)} \chi_{(m_1,m_1+m_2)} ,
\eq
e.g. $\chi_{(m_1,m_1+m_2)}$ does not anticommute with the Dirac matrices $\gamma^\mu_{(m_1+m_2)}$ for $\mu \ge 2 m_1$.
Instead we can show that for $\mu \ge 2 m_1$ we have
\bq
\left[ \gamma^\mu_{(m_1+m_2)}, \chi_{(m_1,m_1+m_2)} \right] & = & 0.
\eq
We will come back to this point in the definition of the HV-like scheme.\\
\\
In supersymmetric theories we have to deal with Majorana spinors. We therefore consider shortly
subtleties of real Clifford algebras:
Up to now we have implicitly assumed that the spinors form a complex vector space.
This is appropriate if one considers Dirac spinors only.
If one imposes a reality condition by considering Majorana spinors, one is led to the study
of real or purely imaginary Clifford algebras instead of complex Clifford algebras.
The classification of complex Clifford algebras is quite trivial, whereas the classification of
real Clifford algebras is a little bit more subtle \cite{atiyah,coquereaux}.
There are basically two types of complex Clifford algebras, 
depending on whether the underlying vector space
has even or odd dimensions. Since we restrict ourselves here to vector spaces of even
dimensions, no complications occur and the Clifford algebra over a vector space of dimension $(2m)$ is isomorphic
to the matrix algebra of dimension $2^m$ over the complex numbers:
\bq
M(2^m,{\Bbb C})
\eq
Real Clifford algebras are classified with a period of 8. Again we will restrict ourselves to vector spaces
of even dimensions. We denote the dimension of the underlying vector space by $(2m)$. We also will need
the signature of the metric of the underlying vectorspace, defined as
\bq
s & = & p - q,
\eq
where the metric $g^{\mu\nu}$ has $p$ plus signs and $q$ minus signs. The structure of the real Clifford algebras
is then given by
\bq
\begin{array}{c|c}
\hline 
(p-q) \; \mbox{mod}\;8 & \mbox{Clifford algebra} \\
\hline
0 & M(2^m,{\Bbb R}) \\
2 & M(2^m,{\Bbb R}) \\
4 & M(2^{m-1}, {\Bbb H}) \\
6 & M(2^{m-1}, {\Bbb H}) \\
\end{array}
\eq
where ${\Bbb H}$ denotes the field of quaternions and $M(n,{\Bbb F})$ denotes the matrix algebra of dimension $n$
over the field ${\Bbb F}$.
Real representations or Majorana spinors exist therefore if $(p-q) \; \mbox{mod}\;8 = 0,2$.
For $(p-q) \; \mbox{mod}\;8 = 0$ the Majorana spinors can be further reduced to Majorana-Weyl spinors.
It can be shown that for massless fermions and $(p-q) \; \mbox{mod}\;8 = 0,6$ a purely imaginary (or 
pseudo-Majorana) representation exists.
We further note that as far as real Clifford algebras are concerned the two choices for the metric
$g^{\mu\nu} = \mbox{diag}\;(1,-1,-1,-1)$ and
$g^{\mu\nu} = \mbox{diag}\;(-1,1,1,1)$ are not equivalent.
We regulate loop integrals involving real Clifford algebras as follows: We consider integration over
\bq
X_1 + X_2,
\eq
where $X_1,X_2 \in C(G)$. $X_1$ has the signature dictated by the problem under consideration, whereas $X_2$
serves to regulate the loop. We may assume that $X_2$ is constructed out of vectorspaces $V_i$ of zero
signature. This does not change the overall signature of $X_1+X_2$.\\
\\
With all necessary tools in hand, we can now address the question how to define spinors in $d$ dimensions.
Put more formally
we consider the problem, how to define spinors over $X_1 + X_2$, given two vectorspaces $X_1$ and $X_2$ of dimensions $(2 m_1)$ and $(2 m_2)$,
respectively.
In the following we denote by $A_{(m)}$ the Clifford algebra corresponding to a vectorspace $X$ of dimension $(2 m)$.
Let us first show why the naive approach to this problem would fail.
The naive approach takes $A_{(m_1+m_2)}$ as the Clifford algebra of $X_1+X_2$.
The continuation
from $(2 m_1)$ to $(2m_1+2m_2)$ dimensions is described by a based linear map $\rho$ from the algebra
$A_{(m_1)}$ to the algebra $A_{(m_1+m_2)}$:
\bq
\rho & : & A_{(m_1)} \rightarrow A_{(m_1+m_2)} 
\eq
The word ``based'' means that the identity of $A_{(m_1)}$ is mapped to the identity of $A_{(m_1+m_2)}$.
The naive scheme is given by
\bq
 \rho\left((pq)_{(m_1)} 1_{(m_1)} \right) & = & (pq)_{(m_1+m_2)} 1_{(m_1+m_2)}, \\
 \rho\left( p\!\!\!/_{(m_1)} \right) & = & p\!\!\!/_{(m_1+m_2)}, \\
 \rho\left( \chi_{(m_1)} \right) & = & \chi_{(m_1+m_2)}.
\eq
We introduce further the bilinear map
\bq
\omega : A_{(m_1)} \times A_{(m_1)} \rightarrow A_{(m_1+m_2)}, 
\eq
defined by
\bq
\omega\left(a_{(m_1)}, b_{(m_1)}\right) = \rho \left( a_{(m_1)} b_{(m_1)}
+ b_{(m_1)} a_{(m_1)} \right)
- \rho \left( a_{(m_1)} \right) \rho \left( b_{(m_1)} \right)
- \rho \left( b_{(m_1)} \right) \rho \left( a_{(m_1)} \right), \nonumber \\
\eq
where $a_{(m_1)}, b_{(m_1)} \in A_{(m_1)}$.
Basically $\omega$ measures if the map $\rho$ preserves the anticommutation relations. 
Within the context of non-commutative geometry the map $\omega$ is also called the curvature of the map
$\rho$.
The naive
scheme considered here preserves the algebraic structure:
\bq
\omega\left( p\!\!\!/_{(m_1)}, q\!\!\!/_{(m_1)}\right) = 0, \\
\omega\left( p\!\!\!/_{(m_1)}, \chi_{(m_1)} \right) = 0
\eq
However, for $m_1=2$ (corresponding to four dimensions) and $m_2 > 0$, we find that $\chi_{(m_1+m_2)}$ is a product
of more than four Dirac matrices, and therefore the trace of four Dirac matrices $\gamma^\mu_{(m_1+m_2)}$ with
$\chi_{(m_1+m_2)}$ vanishes in $A_{(m_1+m_2)}$. Therefore we would get for tree-level amplitudes
different results in the regularized theory (where the trace of four Dirac matrices with one $\chi_{(m_1+m_2)}$
vanishes) and  in the original theory (where this trace does not vanish). We conclude
that the naive scheme is inconsistent.\\
\\
Next we consider a prescription defined by
\bq
\rho\left((pq)_{(m_1)} 1_{(m_1)} \right) & = & (pq)_{(m_1+m_2)} 1_{(m_1+m_2)}, \\
\rho\left( p\!\!\!/_{(m_1)} \right) & = & p\!\!\!/_{(m_1+m_2)}, \\
\rho\left( \chi_{(m_1)} \right) & = & \chi_{(m_1,m_1+m_2)} = - (-i)^{m_1-1} \gamma^0_{(m_1+m_2)} ... \gamma^{2m_1-1}_{(m_1+m_2)}.
\eq
This corresponds to the 't Hooft-Veltman scheme. The Dirac matrices are continued to $2(m_1+m_2)$ dimensions,
whereas for ``$\gamma_5$'' we take the representation of $\chi_{(m_1)}$ in $2(m_1+m_2)$ dimensions.
Now
\bq
\omega\left( p\!\!\!/_{(m_1)}, q\!\!\!/_{(m_1)}\right) & = & 0, \\
\omega\left( p\!\!\!/_{(m_1)}, \chi_{(m_1)} \right) & = & - 2 \sum\limits_{\mu=2m_1}^{2(m_1+m_2)-1} p_\mu^{(m_1+m_2)}
\gamma^\mu_{(m_1+m_2)} \rho\left( \chi_{(m_1)} \right) \neq 0.
\eq
In simple terms this means that the anti-commutation relations of ``$\gamma_\mu$'' with ``$\gamma_\nu$''
is preserved during regularization, whereas the anti-commutation relation of ``$\gamma_\mu$'' with
``$\gamma_5$'' is not preserved.
In order to specify the scheme completely we have to define the continuation of 
$(\gamma^\mu)_{i_1i_2} (\gamma_\mu)_{i_3 i_4}$, where the Dirac matrices belong to different
fermion lines and the index $\mu$ is summed over from 0 to $(2m_1-1)$.
We extend $\rho$ to the tensor algebra and set
\bq
\rho \left( \sum\limits_{\mu=0}^{2m_1-1} \gamma^\mu_{(m_1)} \otimes \gamma_\mu^{(m_1)} \right)
& = & \sum\limits_{\mu=0}^{2m_1+2m_2-1} \gamma^\mu_{(m_1+m_2)} \otimes \gamma_\mu^{(m_1+m_2)}.
\eq
Finally we have to specify the trace. It is common practice to normalize the trace to its
four-dimensional value:
\bq
\mbox{Tr} & : & A_{(m_1+m_2)} \rightarrow {\Bbb C}, \nonumber \\
& & a\!\!\!/_{(m_1+m_2)} \rightarrow \frac{2^{m_1}}{\mbox{tr}\; 1_{(m_1+m_2)}} \mbox{tr}\; a\!\!\!/_{(m_1+m_2)},
\eq
where $\mbox{tr}$ denotes the ordinary trace of $2^{m_1+m_2} \times 2^{m_1+m_2}$-matrices.\\
\\
Finally we consider a scheme where the Dirac algebra of $X_1+X_2$ is given by $A_{(m_1)} + A_{(m_2)}$.
We will denote elements of $A_{(m_1)}+A_{(m_2)}$ by matrix notation:
\bq
\left( \begin{array}{cc}
a\!\!\!/_{(m_1)} & 0 \\
0 & b\!\!\!/_{(m_2)} \\
\end{array}
\right) ,
\eq
where $a\!\!\!/_{(m_1)} \in A_{(m_1)}$ and $b\!\!\!/_{(m_2)} \in A_{(m_2)}$.
The map $\rho$ goes now from $A_{(m_1)}$ to $A_{(m_1)} + A_{(m_2)}$
\bq
\rho & : & A_{(m_1)} \rightarrow A_{(m_1)} + A_{(m_2)}
\eq
and is specified by
\bq
\rho\left((pq)_{(m_1)} 1_{(m_1)} \right) & = & (pq)_{(m_1+m_2)} \left( \begin{array}{cc}
                                                      1_{(m_1)} & 0 \\
                                                      0 & 1_{(m_2)} \\
                                               \end{array} \right), \\
\rho\left( p\!\!\!/_{(m_1)} \right) & = & \left( \begin{array}{cc}
                                                      p\!\!\!/_{(m_1)} & 0 \\
                                                      0 & p\!\!\!/_{(m_2)} \\
                                               \end{array} \right), \\
\rho\left( \chi_{(m_1)} \right) & = & \left( \begin{array}{cc}
                                                      \chi_{(m_1)} & 0 \\
                                                      0 & \chi_{(m_2)} \\
                                               \end{array} \right).
\eq
With these definitions we find for the map $\omega$
\bq
\omega & : & A_{(m_1)} \times A_{(m_1)} \rightarrow A_{(m_1)} + A_{(m_2)},
\eq
\bq
\omega\left( p\!\!\!/_{(m_1)}, q\!\!\!/_{(m_1)}\right) & = & \left( \begin{array}{cc}
 (pq)_{(m_2)} & 0 \\
 0 & (pq)_{(m_1)} \\
\end{array} \right), \\
\omega\left( p\!\!\!/_{(m_1)}, \chi_{(m_1)} \right) & = & 0.
\eq
In simple terms this means that now the anti-commutation relation of ``$\gamma_\mu$'' with ``$\gamma_5$''
is preserved, whereas the anti-commutation relation of ``$\gamma_\mu$'' with ``$\gamma_\nu$'' is not.
We further have
\bq
\rho \left( \sum\limits_{\mu=0}^{2m_1-1} \gamma^\mu_{(m_1)} \otimes \gamma_\mu^{(m_1)} \right)
& = & \sum\limits_{\mu=0}^{2m_1-1} 
\left( \begin{array}{cc} \gamma_\mu^{(m_1)} & 0 \\ 0 & 0 \end{array} \right) \otimes
\left( \begin{array}{cc} \gamma^\mu_{(m_1)} & 0 \\ 0 & 0 \end{array} \right) \nonumber \\
& & +  \sum\limits_{\mu=0}^{2m_2-1} 
\left( \begin{array}{cc} 0 & 0 \\ 0 & \gamma_\mu^{(m_2)} \end{array} \right) \otimes
\left( \begin{array}{cc} 0 & 0 \\ 0 & \gamma^\mu_{(m_2)} \end{array} \right),
\eq
as well as
\bq
\mbox{Tr} & : & A_{(m_1)} + A_{(m_2)} \rightarrow {\Bbb C} \nonumber \\
& & a\!\!\!/ \rightarrow \mbox{tr} \left(
\begin{array}{cc} a\!\!\!/_{(m_1)} & 0 \\ 0 & a\!\!\!/_{(m_2)} \\ \end{array} \right).
\eq
In this scheme external particles may be taken to lie within $X_1$ and act therefore
as projectors onto $A_{(m_1)}$. 
In the next section we will show that with a suitable construction for the gauge group the entire
Dirac algebra may be performed in $A_{(m_1)}$.

\section{The gauge group}

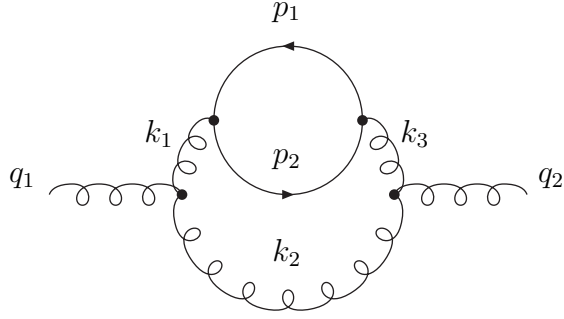
\begin{figure}
\begin{center}
\begin{picture}(300,100)(0,0)
\Gluon(60,50)(110,50){4}{3}
\Gluon(190,50)(240,50){4}{3}
\Vertex(110,50){2}
\Vertex(190,50){2}
\GlueArc(150,50)(40,180,360){4}{7}
\GlueArc(150,50)(40,135,180){4}{2}
\GlueArc(150,50)(40,0,45){4}{2}
\Vertex(178,78){2}
\Vertex(122,78){2}
\ArrowArc(150,78)(28,0,180)
\ArrowArc(150,78)(28,180,360)
\Text(150,118)[b]{$p_1$}
\Text(150,62)[b]{$p_2$}
\Text(150,25)[b]{$k_2$}
\Text(107,70)[br]{$k_1$}
\Text(193,70)[bl]{$k_3$}
\Text(50,55)[b]{$q_1$}
\Text(250,55)[b]{$q_2$}
\end{picture}
\caption{\label{twoloop} A two-loop correction to the gluon propagator}
\end{center}
\end{figure}

Up to now we have constructed ``spaces'', which are in some way the continuation of Minkowski space
and we have defined spinors over these spaces.
We now want to specify how the gauge group is related to this construction.
In this section we will take the colour group $SU(3)$ as an example. The generalization to other gauge groups
is straightforward. (Electroweak and supersymmetric theories are considered explicitly in the next section.)
In the case of the HV-like scheme there is not much choice: We considered the vectorspace
$X_1+X_2$ and defined the spinors as a spinor bundle over $(X_1+X_2)$. Since the quarks are
supposed to be in the fundamental representation of the gauge group, quarks are described
by a vector bundle where the fibre is a vectorspace in the fundamental representation of the
gauge group and the base space is given by the spinor bundle constructed above.
Similarly gluons are described by a vector bundle, whose fibre is a vector space in the adjoint
representation of the gauge group and whose base space is $X_1+X_2$.
Since the Feynman rules for the non-abelian vertices involve the metric tensor, we still have to specify
the trace over the metric tensor. It is common practice in the HV-scheme to take the trace to be
\bq
g^\mu_{(m_1+m_2)\mu} & = & \mbox{rank}\;(X_1+X_2) .
\eq
With $\mbox{rank}\;(X_1) = 4$ and $\mbox{rank}\;(X_2)= -2 \eps$ this gives just
$g^\mu_{(m_1+m_2)\mu} = 4 - 2 \eps$.\\
\\
In the case of the four-dimensional scheme there is however a choice: We have first defined a spinor
bundle over $X_1$ and a different one over $X_2$.
We have therefore some freedom in defining the transformations of spinors under gauge transformations.
Let us denote a spinor over $X_1$ by $\psi_{(m_1)}$ and a spinor over $X_2$ by $\psi_{(m_2)}$.
Various possibilities are:
\begin{enumerate}
\item The first possibility is the case already discussed for the HV-like scheme:
$\psi_{(m_1)}$ and $\psi_{(m_2)}$ are both in the fundamental representation of the gauge
group and a gauge transformation transforms simultaneously $\psi_{(m_1)}$ and $\psi_{(m_2)}$.
\item $\psi_{(m_1)}$ and $\psi_{(m_2)}$ are both in the fundamental representation of the gauge
group, but there is one copy of the gauge group, which acts only on $\psi_{(m_1)}$, but not
on $\psi_{(m_2)}$, as well as a second copy of the gauge group, which acts on $\psi_{(m_2)}$ but not
on $\psi_{(m_1)}$.
\item Only $\psi_{(m_1)}$ transforms as the fundamental representation of the gauge group, whereas
$\psi_{(m_2)}$ is a singlet. 
\end{enumerate}
In appendix C we have rephrased the various prescriptions in a more mathematical way.
We will take prescription 2 as our definition. We discuss the differences between the various options 
by an example given in fig. \ref{twoloop}.
We will regulate the loops by considering $X_1+X_2+X_3$. $X_3$ is used to regulate the fermion loop.
The momenta $p_1$ and $p_2$ live in $X_1+X_2+X_3$, and $X_1+X_2$ is supposed to be the space of external
momenta with respect to the fermion loop.
$X_2$ is used to regulate the second loop, $k_1$, $k_2$ and $k_3$ live in $X_1+X_2$ and $X_1$ is supposed
to be the space of external momenta for this second loop. Since there are no further loops left, 
$X_1$ is identified with the physical Minkowski space.
The naive Feynman rule for the three-gluon vertex in the unregularized theory reads
\bq
g f^{abc} \left[ (k_3-k_2)_\mu g_{\nu\lambda} + ( k_1 - k_3)_\nu g_{\lambda\mu}
+(k_2-k_1)_\lambda g_{\mu\nu} \right].
\eq
If we choose prescription 1 we obtain the following Feynman rule for the three-gluon vertex in the 
regularized theory:
\bq
& & g f^{abc} \left[ (k^{(m_1+m_2)}_3-k^{(m_1+m_2)}_2)_\mu g^{(m_1+m_2)}_{\nu\lambda} + ( k^{(m_1+m_2)}_1 - k^{(m_1+m_2)}_3)_\nu g^{(m_1+m_2)}_{\lambda\mu} \right. \nonumber \\
& & \;\;\; \left.+(k^{(m_1+m_2)}_2-k^{(m_1+m_2)}_1)_\lambda g^{(m_1+m_2)}_{\mu\nu} \right].
\eq
We will foccus on the contribution where at each three-gluon vertex the metric tensor contracts
the two gluons in the loop. If we follow prescription 1 we obtain the following contribution:
\bq
& & i g^4 f^{acd} f^{bde} \mbox{Tr}\; \left( T^c T^e \right) \int \frac{d^dk_1}{(2 \pi)^d} \int \frac{d^dp_1}{(2 \pi)^d}
\frac{1}{k_1^2 k_2^2 k_3^2}
\frac{1}{p_1^2 p_2^2} \nonumber \\
& & \eps_1^{(m_1)} \cdot (k_1^{(m_1+m_2)}+k_2^{(m_1+m_2)}) \eps_2^{(m_1)} \cdot (k_2^{(m_1+m_2)}+k_3^{(m_1+m_2)}) \nonumber \\
& & \mbox{Tr}\; \left( p\!\!\!/_1^{(m_1+m_2+m_3)} \gamma_\mu^{(m_1+m_2)} p\!\!\!/_2^{(m_1+m_2+m_3)} \gamma^\mu_{(m_1+m_2)} \right)
\eq
Here $\eps_1^{(m_1)}$ and $\eps_2^{(m_2)}$ denote the polarization vectors of the external gluons.
We see that the Dirac algebra of the fermion loop is only projected onto $X_1+X_2$, but not onto
$X_1$.\\
\\
We now consider precription 2: Here each vectorspace $X_1$, $X_2$ and $X_3$ has its on copy of the gauge
group. A gauge transformation in $X_1$ depends only on the coordinates of $X_1$ and does not affect
$X_2$ or $X_3$. Similarly, a gauge transformation in $X_2$ depends only on the coordinates of $X_2$ and
does not affect $X_1$ or $X_3$.
The gluon propagator on $X_1+X_2$ is therefore
\bq
\frac{-i}{k^2_{(m_1+m_2)}}
\left\{ \delta^{ab}_{(m_1)} \left[ g^{\mu\nu}_{(m_1)} - ( 1 - \xi_{(m_1)}) \frac{k^\mu_{(m_1)} k^\nu_{(m_1)}}{k^2_{(m_1+m_2)}} \right]
+ \delta^{ab}_{(m_2)} \left[ g^{\mu\nu}_{(m_2)} - ( 1 - \xi_{(m_2)}) \frac{k^\mu_{(m_2)} k^\nu_{(m_2)}}{k^2_{(m_1+m_2)}} \right] \right\}. \nonumber \\
\eq
The gluon propagator consists of two pieces, one piece describes the propagation in $X_1$, the other
one describes the propagation in $X_2$. The colour factors $\delta^{ab}_{(m_1)}$ and $\delta^{ab}_{(m_2)}$
ensure that the two pieces do not mix. $\xi_{(m_1)}$ and $\xi_{(m_2)}$ are the gauge-fixing parameters in $X_1$
and in $X_2$, respectively.
\\
The three-gluon vertex is given by
\bq
& & g \left\{ f^{abc}_{(m_1)}  \left[ (k_{3}^{(m_1)} - k_{2}^{(m_1)})_{\mu} g^{(m_1)}_{\nu \lambda} 
               + (k^{(m_1)}_{1} - k^{(m_1)}_{3})_{\nu} g^{(m_1)}_{\lambda \mu} 
         + (k^{(m_1)}_{2} - k^{(m_1)}_{1})_{\lambda} g^{(m_1)}_{\mu \nu} \right] \right. \nonumber \\
& & + \left. f^{abc}_{(m_2)}  \left[ (k_{3}^{(m_2)} - k_{2}^{(m_2)})_{\mu} g^{(m_2)}_{\nu \lambda} 
               + (k^{(m_2)}_{1} - k^{(m_2)}_{3})_{\nu} g^{(m_2)}_{\lambda \mu} 
         + (k^{(m_2)}_{2} - k^{(m_2)}_{1})_{\lambda} g^{(m_2)}_{\mu \nu} \right] \right\}. \nonumber \\
\eq
For the two-loop example considered above we obtain with prescription 2:
\bq
& & i g^4 f^{acd}_{(m_1)} f^{bde}_{(m_1)} \mbox{Tr}\; \left( T^c_{(m_1)} T^e_{(m_1)} \right) \int \frac{d^dk_1}{(2 \pi)^d} \int \frac{d^dp_1}{(2 \pi)^d}
\frac{1}{k_1^2 k_2^2 k_3^2}
\frac{1}{p_1^2 p_2^2} \nonumber \\
& & \eps_1^{(m_1)} \cdot (k_1^{(m_1)}+k_2^{(m_1)}) \eps_2^{(m_1)} \cdot (k_2^{(m_1)}+k_3^{(m_1)}) \nonumber \\
& & \mbox{Tr}\; \left( p\!\!\!/_1^{(m_1)} \gamma_\mu^{(m_1)} p\!\!\!/_2^{(m_1)} \gamma^\mu_{(m_1)} \right)
\eq
The colour indices ensure now the projection onto $X_1$.
We see that the Dirac algebra can be performed entirely in four dimensions.
Intuitively we have the following picture: Particles in $X_1$ carry colour charges of $SU(3)_{(m_1)}$ but
not of $SU(3)_{(m_2)}$. On the other hand hypothetical particles in $X_2$ carry colour charges of $SU(3)_{(m_2)}$
but not of $SU(3)_{(m_1)}$. Therefore the two sectors cannot couple to each other. If we now require
that all external particles live in $X_1$, it follows that only the $SU(3)_{(m_1)}$-part of each
internal particle contributes.
We complete the list of Feynman rules for QCD following presription 2:
The quark propagator reads
\bq
\frac{i}{p_{(m_1+m_2)}^2-m^2} \left(\begin{array}{cc}
\delta^{(m_1)}_{ij} \left( p\!\!\!/_{(m_1)} +m \right)& 0 \\
0 & \delta^{(m_2)}_{ij} \left( p\!\!\!/_{(m_2)} +m \right)\\
\end{array}
\right).
\eq
The ghost propagator is given by
\bq
\frac{i}{k^2_{(m_1+m_2)}} \left( \delta^{ab}_{(m_1)} + \delta^{ab}_{(m_2)} \right) .
\eq
The quark-gluon vertex is given by
\bq
i g \left( \begin{array}{cc}
\gamma_{\mu}^{(m_1)} T^a_{(m_1) ij} & 0 \\
0 & \gamma_{\mu}^{(m_2)} T^a_{(m_2) ij} \\
\end{array} \right).
\eq
The four-gluon-vertex reads
\bq
& & - i g^{2} \left\{ \left[ f^{abe}_{(m_1)} f^{ecd}_{(m_1)} 
\left( g^{(m_1)}_{\mu \lambda} g^{(m_1)}_{\nu \rho} - g^{(m_1)}_{\mu \rho} g^{(m_1)}_{\nu \lambda} \right) 
+ f_{(m_1)}^{ace} f_{(m_1)}^{ebd} \left( g^{(m_1)}_{\mu \nu} g^{(m_1)}_{\lambda \rho} - g^{(m_1)}_{\mu \rho} g^{(m_1)}_{\lambda \nu} \right) \right. \right. \nonumber \\
& & \left. \left. + f_{(m_1)}^{ade} f_{(m_1)}^{ecb} \left( g^{(m_1)}_{\mu \nu} g^{(m_1)}_{\lambda \rho} - g^{(m_1)}_{\mu \lambda} g^{(m_1)}_{\nu \rho} \right) \right] \right. \nonumber \\
& & + \left. \left[ f^{abe}_{(m_2)} f^{ecd}_{(m_2)} 
\left( g^{(m_2)}_{\mu \lambda} g^{(m_2)}_{\nu \rho} - g^{(m_2)}_{\mu \rho} g^{(m_2)}_{\nu \lambda} \right) 
+ f_{(m_2)}^{ace} f_{(m_2)}^{ebd} \left( g^{(m_2)}_{\mu \nu} g^{(m_2)}_{\lambda \rho} - g^{(m_2)}_{\mu \rho} g^{(m_2)}_{\lambda \nu} \right) \right. \right. \nonumber \\
& & \left. \left. + f_{(m_2)}^{ade} f_{(m_2)}^{ecb} \left( g^{(m_2)}_{\mu \nu} g^{(m_2)}_{\lambda \rho} - g^{(m_2)}_{\mu \lambda} g^{(m_2)}_{\nu \rho} \right) \right] \right\}.
\eq
Finally the gluon-ghost vertex is given by
\bq
g \left( f^{abc}_{(m_1)} k^{(m_1)}_{\mu} + f^{abc}_{(m_2)} k^{(m_2)}_{\mu} \right).
\eq
With these rules we can show that given a connected Feynman diagram, where all external particles
are in the physical Minkowski space, that the Dirac algebra may be performed in four dimensions.
The proof relies on the observation that no propagator nor any vertex induces a mixing between components
of $X_1$ and $X_2$. \\
\\
We do not consider prescription 3 here, since it will lead for external particles in $X_1$ to the same results
as prescription 2.

\section{Electroweak and supersymmetric theories}

In the previous section we have defined the four-dimensional scheme for QCD. We have chosen QCD as an example
to explain the basic ideas. The construction is however more general and can be extended without any
problems to other theories.
The first extension would be to include the electroweak sector, with different couplings to left- and
right handed fermions.
Using the projection operators $P^+_{(m_1)}$ and $P^-_{(m_1)}$ from eq. (\ref{projectors}) we may split the spinor
bundle over $X_1$ in left- and righthanded spinors.
We then introduce the electroweak gauge group $U(1)_Y \times SU(2)_L$ and couple it in the standard way
to the left- and right handed spinors.
We proceed similar for the spinor bundle over $X_2$: We first split it, using $P^+_{(m_2)}$ and $P^-_{(m_2)}$,
and introduce a second copy of $U(1)_Y \times SU(2)_L$, which is then coupled to the spinors over $X_2$.
Again we have used two copies of the gauge group.\\
\\
The recipe works also for supersymmetric theories: We introduce spinor charges
$Q^{(m_1)}_\alpha$ and $\bar{Q}^{(m_1)}_{\dot{\alpha}}$, satisfying 
the supersymmetry algebra
\bq
\left\{ Q^{(m_1)}_\alpha, \bar{Q}^{(m_1)}_{\dot{\alpha}} \right\} & = & 
2 \sigma^{(m_1)\mu}_{\alpha \dot{\alpha}} P^{(m_1)}_\mu, 
\eq
which together with the generators of boosts and rotations $M^{\mu\nu}_{(m_1)}$ as well as with the
generators of translations $P_\mu^{(m_1)}$
form the super-Poincar\'e algebra in $2m_1$ dimensions. Similarly we may introduce the super-Poincar\'e algebra 
for $X_2$, generated by $Q_\alpha^{(m_2)}$, $\bar{Q}_{\dot{\alpha}}^{(m_2)}$, $M^{\mu\nu}_{(m_2)}$ and $P_\mu^{(m_2)}$.
In general the super-Poincar\'e algebra of $X_1$ will not be isomorph to the super-Poincar\'e algebra of $X_2$,
unless $m_1 = m_2$ (e.g. $X_1$ and $X_2$ have the same dimension).
In the case of supersymmetric theories it is helpful to clarify the r\^ole of various ``Lorentz groups'':
Our starting point was a set of vector spaces, where each element was a representations of the physical Lorentz group
$SO(1,3)$. Therefore $X_1$ and $X_2$ are also representations of $SO(1,3)$. We used this fact to show that the
HV-like and the four-dimensional scheme can be formulated in a Lorentz invariant way. 
However the construction of the regularization scheme does not rely on the fact that all vectorspaces are
representation of $SO(1,3)$, any set of even-dimensional vector spaces would do.
Now we want to formulate a four-dimensional scheme for supersymmetric theories.
We forget about the action of $SO(1,3)$ on $X_1$ and $X_2$ and introduce the super-Poincar\'e algebra for $X_1$ 
and $X_2$ separately.
This is in accordance with our general observation that practical calculations simplify if each
sector $X_1$ and $X_2$ has its own symmetry group.
The generators $M^{\mu\nu}_{(m_1)}$ and $M^{\mu\nu}_{(m_2)}$ generate the groups $SO(1,2m_1-1)$ and $SO(p,2m_2-p)$ 
(where $p$ is an integer depending on the signature of the metric in $X_2$).
The indices of $M^{\mu\nu}_{(m_1)}$ take values in $0 \le \mu,\nu \le (2m_1-1)$, whereas the indices of
$M^{\mu\nu}_{(m_2)}$ take values in $0 \le \mu,\nu \le (2m_2-1)$.\\
\\
The extension of the four-dimensional scheme from supersymmetric theories to supergravity is straightforward.

\section{The scheme in practice}

In this section we will show how the scheme may be used in practice. We will foccus on one-loop calculations
as an example. The generalization to higher loops is straightforward.
It is convenient to consider the integration over
\bq
X_0 + X_1,
\eq
where $X_0$ is the representative in $C(G)$ of the physical Minkowski space and $X_1$ is an
element of $C(G)$ of rank $(-2 \eps)$.
We may assume that we have a representative of $X_1$, which is a vector space of dimension
$2 m_\eps$.
It is important to note that $2 m_\eps$ is always an integer number. 
The distinction between the dimension (e.g. $2 m_\eps$ for $X_1$) and the rank (e.g. $-2 \eps$ for $X_1$) is a fundamental
property of our scheme.
The representative
of $X_0+X_1$ is therefore a finite-dimensional vectorspace, whose dimension is by construction
even and always greater than four.
It is the clear separation between the rank ($4-2\eps$) and the dimension of a representative ($4+2m_\eps$)
which avoids inconsistencies inherent in other approaches.
In the previous section we have introduced two possible continuations of the Dirac algebra.
In the HV-like scheme the Dirac matrices become matrices of
dimension $2^{2+m_\eps} \times 2^{2+m_\eps}$. Practical calculations proceed as in the
original formulation of the 't Hooft-Veltman scheme.
In our formulation of the HV-like scheme there are no additional simplifications nor any additional
complications as compared to the standard formulation of the 't Hooft-Veltman scheme.\\
\\
It is the aim of this paper to recommend the four-dimensional scheme for future calculations.
In the four-dimensional scheme the Dirac algebra is continued to $A_{(2)}+A_{(m_\eps)}$ and the fermion
propagator becomes a matrix of size
$ (4 + 2^{m_\eps}) \times (4 + 2^{m_\eps})$.
Since all external particles lie in the physical Minkowski space $X_0$, they act effectively as
projection operators onto $A_{(2)}$. We have shown in the previous section that with a suitable
construction for the gauge group, the entire Dirac algebra may be performed in four dimensions.
This simplifies practical calculations considerably. Assuming that all external particles
lie in the physical Minkowski space and that all Feynman diagrams are connected we may
write down a simplified list of Feynman rules for QCD in the four-dimensional scheme:
The gluon propagator in a covariant gauge is given by
\bq
\frac{-i}{k^2_{(2+m_\eps)}}
\delta^{ab}_{(2)} \left[ g^{\mu\nu}_{(2)} - ( 1 - \xi_{(2)}) \frac{k^\mu_{(2)} k^\nu_{(2)}}{k^2_{(2+m_\eps)}} \right].
\eq
The quark propagator reads
\bq
\frac{i}{p_{(2+m_\eps)}^2-m^2} 
\delta^{(2)}_{ij} \left( p\!\!\!/_{(2)} +m \right).
\eq
The ghost propagator is given by
\bq
\frac{i}{k^2_{(2+m_\eps)}} \delta^{ab}_{(2)} .
\eq
The quark-gluon vertex is given by
\bq
i g 
\gamma_{\mu}^{(2)} T^a_{(2) ij} .
\eq
The three-gluon vertex is given by
\bq
& & g f^{abc}_{(2)}  \left[ (k_{3}^{(2)} - k_{2}^{(2)})_{\mu} g^{(2)}_{\nu \lambda} 
               + (k^{(2)}_{1} - k^{(2)}_{3})_{\nu} g^{(2)}_{\lambda \mu} 
         + (k^{(2)}_{2} - k^{(2)}_{1})_{\lambda} g^{(2)}_{\mu \nu} \right]. 
\eq
The four-gluon-vertex reads
\bq
& & - i g^{2} \left[ f^{abe}_{(2)} f^{ecd}_{(2)} 
\left( g^{(2)}_{\mu \lambda} g^{(2)}_{\nu \rho} - g^{(2)}_{\mu \rho} g^{(2)}_{\nu \lambda} \right) 
+ f_{(2)}^{ace} f_{(2)}^{ebd} \left( g^{(2)}_{\mu \nu} g^{(2)}_{\lambda \rho} - g^{(2)}_{\mu \rho} g^{(2)}_{\lambda \nu} \right) \right. \nonumber \\
& & \left. + f_{(2)}^{ade} f_{(2)}^{ecb} \left( g^{(2)}_{\mu \nu} g^{(2)}_{\lambda \rho} - g^{(2)}_{\mu \lambda} g^{(2)}_{\nu \rho} \right) \right] .
\eq
Finally the gluon-ghost vertex is given by
\bq
g f^{abc}_{(2)} k^{(2)}_{\mu} .
\eq
In simple words, all open indices (Lorentz, spinor and colour indices) are treated as in four dimensions, whereas the denominators of the
propagators are continued to ``$d$ dimensions''.
The fact that the entire algebra may be performed in four dimensions leads to large simplifications
in practical calculations.
It allows the application of spinor helicity methods. Furthermore the four-dimensional
Schouten- and Fierz identies may be used for simplifications.
The Fierz identity reads in the bra-ket notation \footnote{There are two notations in the literature for Weyl spinors:
the bra-ket notation as well as the notation with dotted and undotted indices. The relation between the two
notations is: $|p+\rangle =p_A$, $|p-\rangle =p^{\dot{A}}$, $\langle q+|=q_{\dot{B}}$ and $\langle q-|=q^B$.}:
\bq
\langle p^{(2)}_1+ | \gamma^{(2)}_{\mu} | p^{(2)}_2+ \rangle \langle p^{(2)}_3- | \gamma_{(2)}^{\mu} | p^{(2)}_4- \rangle & = &
2 [p^{(2)}_1 p^{(2)}_4] \langle p^{(2)}_3 p^{(2)}_2 \rangle
\eq 
The Schouten identity for two-component spinors is given by
\bq
\langle p^{(2)}_1 p^{(2)}_2 \rangle \langle p^{(2)}_3 p^{(2)}_4 \rangle & = & \langle p^{(2)}_1 p^{(2)}_4 \rangle \langle p^{(2)}_3 p^{(2)}_2 \rangle 
 + \langle p^{(2)}_1 p^{(2)}_3 \rangle \langle p^{(2)}_2 p^{(2)}_4 \rangle, \nonumber \\
\left[p^{(2)}_1 p^{(2)}_2\right] [p^{(2)}_3 p^{(2)}_4] & = & [p^{(2)}_1 p^{(2)}_4] [p^{(2)}_3 p^{(2)}_2] + [p^{(2)}_1 p^{(2)}_3] [p^{(2)}_2 p^{(2)}_4].
\eq
where the spinor products are defined in terms of the two-component Weyl spinors $p_{(2)}^A$ and $p^{(2)}_{\dot{A}}$ as
\begin{eqnarray*}
\langle p^{(2)} q^{(2)} \rangle & = & p_{(2)}^A q^{(2)}_A, \\
\left[ p^{(2)} q^{(2)} \right] & = & p^{(2)}_{\dot{A}} q_{(2)}^{\dot{A}}.
\end{eqnarray*}
We may also use the Schouten identity for four-vectors:
\begin{eqnarray}
\eps(q^{(2)}_1,q^{(2)}_2,q^{(2)}_3,q^{(2)}_4) k^{(2)}_\mu & = & \eps(\mu,q^{(2)}_2,q^{(2)}_3,q^{(2)}_4) \left( k^{(2)} \cdot q^{(2)}_1 \right)
+\eps(q^{(2)}_1,\mu,q^{(2)}_3,q^{(2)}_4) \left( k^{(2)} \cdot q^{(2)}_2 \right) \nonumber \\
& & +\eps(q^{(2)}_1,q^{(2)}_2,\mu,q^{(2)}_4) \left( k^{(2)} \cdot q^{(2)}_3 \right) +\eps(q^{(2)}_1,q^{(2)}_2,q^{(2)}_3,\mu) \left( k^{(2)} \cdot q^{(2)}_4 \right), \nonumber \\
\end{eqnarray}
where the notation $\eps(\mu,q^{(2)}_1,q^{(2)}_2,q^{(2)}_3) = 4 i \eps^{(2)}_{\mu\nu\rho\sigma} q^{(2)\nu}_2 q^{(2)\rho}_3 q^{(2)\sigma}_4$ is used.\\
\\
A little bit of care has to be taken for expressions like
\bq
k\!\!\!/_{(2)} k\!\!\!/_{(2)}.
\eq
This gives $(k^{(2)})^2$,e.g. the four-dimensional value, and does not cancel exactly a propagator
$k^2 = (k^{(2)})^2 + (k^{(m_\eps)})^2$.
However this complication occurs at the level of scalar integrals and is most efficiently dealt
with by expressing the additional scalar integrals with powers of $(k^{(m_\eps)})^2$ in the numerator
as integrals in higher dimensions \cite{Bern}:
\bq
\int \frac{d^{4-2\eps}k}{(2 \pi)^{4-2\eps}} \frac{\left(\left(k^{(m_\eps)}\right)^2 \right)^r}{k_0^2 ... k_n^2}
& = & \eps (\eps-1) (\eps-2) ... (\eps-(r-1)) \left( 4 \pi \right)^r
\int \frac{d^{4+2r-2\eps}k}{(2 \pi)^{4+2r-2\eps}} \frac{1}{k_0^2 ... k_n^2} \nonumber \\
\eq
Finally we would like to indicate how the Feynman rules are generalized to two loops and beyond.
Multiloop integrals are regularized by considering integration over
\bq
X_0 + X_1 + ... + X_l,
\eq
where $X_0$ denotes again the physical Minkowski space of dimension $4$ and rank $4$.
$X_1$ regulates the outermost loop and is a vectorspace of dimension $2m_1$ and rank $-2\eps$.
For an $l$-loop integral the remaining $X_j$ (with $1 < j \le l$) serve to regulate the remaining $l-1$
loops. $X_j$ is a vectorspace of dimension $2m_j$ and rank $0$. We have chosen $\mbox{rank}\;X_j=0$ for $j>1$
such that each loop integration (over $X_0+X_1+...+X_l$, $X_0+X_1+...+X_{l-1}$,...,$X_0+X_1$) is always of
rank $4-2\eps$ in accordance with the standard prescription of conventional dimensional regularization.
The Feynman rules for multiloop calculations are straightforward and follow the rule that uncontracted
indices are always in four dimensions, whereas the denominators of the propagators are continued
to ``$d$ dimensions''. For example the quark propagator appearing in the innermost loop of an $l$-loop
diagram would read:
\bq
\frac{i}{p_{(2+m_1+m_2+...+m_{l})}^2-m^2} 
\delta^{(2)}_{ij} \left( p\!\!\!/_{(2)} +m \right)
\eq

\section{Examples}

We now give a few one-loop examples how the scheme is used in practice. We do the calculations both in the
HV-like scheme as well as in the four-dimensional scheme.
We calculate first the triangle anomaly for one axial vector current and two vector currents.
We calculate both the divergence of the axial vector current (AVV-anomaly) and of the vector current (VVA-anomaly).
We then replace the two vector currents by two axial vector currents and repeat the calculation for the
so-called AAA-anomaly.
Finally we consider the Ward identities for the non-singlet axial current and the vector current.
Where no confusion is expected
we will drop the sub- or superscript ${(m)}$, which indicates one half of the dimension
of the corresponding vectorspace. 

\subsection{The singlet axial-vector current and the triangle anomaly}

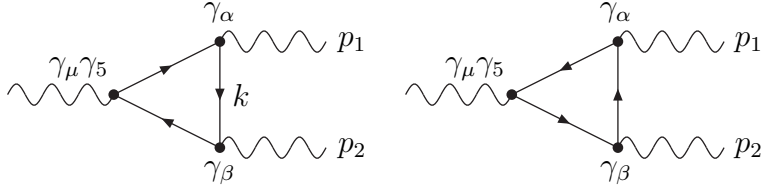
\begin{figure}
\begin{center}
\begin{picture}(300,100)(0,0)
\ArrowLine(30,50)(70,70)
\ArrowLine(70,70)(70,30)
\ArrowLine(70,30)(30,50)
\Vertex(30,50){2}
\Vertex(70,70){2}
\Vertex(70,30){2}
\Photon(-10,50)(30,50){4}{3}
\Photon(70,70)(110,70){4}{3}
\Photon(70,30)(110,30){4}{3}
\Text(28,59)[rb]{$\gamma_\mu \gamma_5$} 
\Text(70,80)[b]{$\gamma_\alpha$}
\Text(70,25)[t]{$\gamma_\beta$}
\Text(75,50)[l]{$k$}
\Text(115,70)[l]{$p_1$}
\Text(115,30)[l]{$p_2$}
\ArrowLine(220,70)(180,50)
\ArrowLine(220,30)(220,70)
\ArrowLine(180,50)(220,30)
\Vertex(180,50){2}
\Vertex(220,70){2}
\Vertex(220,30){2}
\Photon(140,50)(180,50){4}{3}
\Photon(220,70)(260,70){4}{3}
\Photon(220,30)(260,30){4}{3}
\Text(178,59)[rb]{$\gamma_\mu \gamma_5$} 
\Text(220,80)[b]{$\gamma_\alpha$}
\Text(220,25)[t]{$\gamma_\beta$}
\Text(265,70)[l]{$p_1$}
\Text(265,30)[l]{$p_2$}
\end{picture}
\caption{\label{singlet} The triangle graphs for the anomaly}
\end{center}
\end{figure}
For massless quarks we obtain for the sum of the two graphs shown in fig.\ref{singlet}:
\bq
\label{starting_point}
A_{\alpha\beta\mu} & = & -i \int \frac{d^dk}{(2 \pi)^d} \frac{N_{\alpha\beta\mu}}{k^2 \left(k+p_1\right)^2 \left( k -p_2 \right)^2 },
\eq
where
\bq
N_{\alpha\beta\mu} & = & \mbox{Tr}\; \left( k\!\!\!/ + p\!\!\!/_1 \right) \gamma_\alpha k\!\!\!/ \gamma_\beta \left( k\!\!\!/ - p\!\!\!/_2 \right) \gamma_\mu \gamma_5 
- \mbox{Tr}\; \left( k\!\!\!/ - p\!\!\!/_2 \right) \gamma_\beta k\!\!\!/ \gamma_\alpha \left( k\!\!\!/ + p\!\!\!/_1 \right) \gamma_\mu \gamma_5 .
\eq
We will use the notation $k_0=k$, $k_1 = k - p_2$ and $k_2 = k + p_1$ as well as
\bq
\eps(a,b,c,d) & = & 4 i \eps_{\alpha\beta\gamma\delta} a^\alpha b^\beta c^\gamma d^\delta
\eq
for the four-dimensional antisymmetric tensor. It is convenient to calculate the graphs for the kinematical
configuration where $p_1^2$, $p_2^2$ and $(p_1+p_2)^2$ are non-zero. In that case there will be no
infrared divergences, which are not relevant to the discussion of the anomaly.
Contracting $A_{\alpha\beta\mu}$ with $(p_1+p_2)^\mu$ gives the anomaly:
\bq
A^{AVV} & = & \left( p_1 + p_2 \right)^\mu A_{\alpha\beta\mu}
\eq
We first present a calculation of the anomaly along the lines of 't Hooft and Veltman. In the first trace
of $(p_1+p_2)^\mu N_{\alpha\beta\mu}$ we use
\bq
\left( p\!\!\!/_1 + p\!\!\!/_2 \right) \gamma_5 = \left( k\!\!\!/_2 - k\!\!\!/_1 \right) \gamma_5
= - k\!\!\!/_1 \gamma_5 - \gamma_5 k\!\!\!/_2 + 2 k\!\!\!/^{(m_\eps)} \gamma_5.
\eq
For the second line we use
\bq
\left( p\!\!\!/_1 + p\!\!\!/_2 \right) \gamma_5 = \left( k\!\!\!/_2 - k\!\!\!/_1 \right) \gamma_5
=  k\!\!\!/_2 \gamma_5 + \gamma_5 k\!\!\!/_1 - 2 k\!\!\!/^{(m_\eps)} \gamma_5.
\eq
The terms $k\!\!\!/_1 k\!\!\!/_1$ and $k\!\!\!/_2 k\!\!\!/_2$ inside the traces cancel propagators and the
resulting tensor bubble integrals can be shown to vanish after integration.
Therefore the only relevant term is:
\bq
2 \left( \mbox{Tr}\; k\!\!\!/_2 \gamma_\alpha k\!\!\!/_0 \gamma_\beta k\!\!\!/_1 k\!\!\!/^{(m_\eps)} \gamma_5
+ \mbox{Tr}\; k\!\!\!/_1 \gamma_\beta k\!\!\!/_0 \gamma_\alpha k\!\!\!/_2 k\!\!\!/^{(m_\eps)} \gamma_5 \right)
\eq
In the first trace we permute now $k\!\!\!/_2$ to the right and use the fact that traces like
\bq
\mbox{Tr} \; k\!\!\!/_0 \gamma_\beta k\!\!\!/_1 k\!\!\!/^{(m_\eps)} \gamma_5
\eq
vanish. We obtain
\bq
\mbox{Tr}\; k\!\!\!/_2 \gamma_\alpha k\!\!\!/_0 \gamma_\beta k\!\!\!/_1 k\!\!\!/^{(m_\eps)} \gamma_5
& = & \left( k^{(m_\eps)} \right)^2 4 i \eps_{\alpha\lambda\beta\kappa} p_1^\lambda p_2^\kappa.
\eq
A similar result holds for the second trace. We then obtain for the anomaly
\bq
A_{HV}^{AVV} & = & 16 i \eps_{\alpha\lambda\beta\kappa} p_1^\lambda p_2^\kappa \int
\frac{d^dk}{(2 \pi)^d i} \frac{\left(k^{(m_\eps)}\right)^2}{k_0^2 k_1^2 k_2^2} \\
& = & \frac{1}{(4 \pi)^2} 8 i \eps_{\alpha\beta\lambda\kappa} p_1^\lambda p_2^\kappa
\eq
which is the well-known result for the anomaly in the 't Hooft - Veltman scheme \cite{HV}. \\
\\
We now present the calculation of the anomaly in the four-dimensional scheme where the Dirac algebra is performed entirely in
four dimensions. As above we replace $(p_1+p_2)^\mu$ by $(k_2-k_1)^\mu$. Since the Dirac algebra is projected onto
four dimensions we are allowed to write
\bq
\left( p\!\!\!/_1 + p\!\!\!/_2 \right) \gamma_5 = \left( k\!\!\!/_2^{(2)} - k\!\!\!/_1^{(2)} \right) \gamma_5
= - k\!\!\!/_1^{(2)} \gamma_5 - \gamma_5 k\!\!\!/_2^{(2)}, \\
\left( p\!\!\!/_1 + p\!\!\!/_2 \right) \gamma_5 = \left( k\!\!\!/_2^{(2)} - k\!\!\!/_1^{(2)} \right) \gamma_5
=  k\!\!\!/_2^{(2)} \gamma_5 + \gamma_5 k\!\!\!/_1^{(2)} .
\eq
Now
\bq
k\!\!\!/_1^{(2)} k\!\!\!/_1^{(2)} & = & \left( k_1^{(2)} \right)^2 = \left( k_1^{(2+m_\eps)} \right)^2 -
\left( k^{(m_\eps)} \right)^2.
\eq
It is important to note that the four-dimensional $(k_1^{(2)})^2$ does not cancel the propagator exactly, we write
it as the difference of a $2(2+m_\eps)$-dimensional piece $(k_1^{(2+m_\eps)})^2$ and a $2 m_\eps$ dimensional
piece $(k^{(m_\eps)})^2$. The $(k_1^{(2+m_\eps)})^2$ cancels the propagator and the resulting bubble integral
will vanish after integration as it did in the case discussed above.
We are left with the terms proportional to $(k^{(m_\eps)})^2$:
\bq 
\lefteqn{\left( k^{(m_\eps)} \right)^2 \left[ \mbox{Tr}\; k\!\!\!/_2 \gamma_\alpha k\!\!\!/_0 \gamma_\beta \gamma_5
                                    +\mbox{Tr}\; \gamma_\alpha k\!\!\!/_0 \gamma_\beta k\!\!\!/_1 \gamma_5 \right. } & & \nonumber \\
& &   \left.                                 +\mbox{Tr}\; k\!\!\!/_2 \gamma_\beta k\!\!\!/_0 \gamma_\alpha \gamma_5
                                    +\mbox{Tr}\; \gamma_\beta k\!\!\!/_0 \gamma_\alpha k\!\!\!/_2 \gamma_5
\right] \nonumber \\
& = & 2 \left( k^{(m_\eps)} \right)^2 \eps(\alpha,\beta,k_0,p_1+p_2)
\eq
We then obtain for the anomaly
\bq
A_{FD}^{AVV} & = & 2 \int \frac{d^dk}{(2 \pi)^d i} \frac{\eps(\alpha,\beta,k_0,p_1+p_2) \left(k^{(m_\eps)}\right)^2}{k_0^2 k_1^2 k_2^2} .
\eq
This integral is most conveniently done by first introducing Feynman parameters, performing the momentum
integral, expanding in $\eps$ and finally performing the integration over the Feynman parameters. 
The explicit calculation of this integral is given in the appendix.
We obtain
\bq
A_{FD}^{AVV} & = & \frac{1}{3} \frac{1}{(4 \pi)^2} 8 i \eps_{\alpha\beta_\lambda\kappa} p_1^\lambda p_2^\kappa.
\eq
We note that this result differs by a factor of $1/3$ from the result in the 't Hooft-Veltman scheme.\\
\\
We now check the conservation of the vector current. To this aim we contract eq. (\ref{starting_point})
with $p_1^\alpha = k_2^\alpha-k_0^\alpha$:
\bq
A^{VVA} & = & p_1^\alpha A_{\alpha\beta\mu}
\eq
In the HV-like scheme we find:
\bq
A^{VVA}_{HV} & = & \int \frac{d^dk}{(2\pi)^d i} \frac{1}{k_0^2 k_1^2 k_2^2} \left\{
k_2^2 \mbox{Tr}\; k\!\!\!/_0 \gamma_\beta k\!\!\!/_1 \gamma_\mu \gamma_5
-k_0^2 \mbox{Tr}\; k\!\!\!/_2 \gamma_\beta k\!\!\!/_1 \gamma_\mu \gamma_5 \right. \nonumber \\
& & \left. -k_2^2 \mbox{Tr}\; k\!\!\!/_1 \gamma_\beta k\!\!\!/_0 \gamma_\mu \gamma_5
+k_0^2 \mbox{Tr}\; k\!\!\!/_1 \gamma_\beta k\!\!\!/_2 \gamma_\mu \gamma_5 \right\} \nonumber \\
& = & 0,
\eq
since the remaining bubble integrals vanish after integration. In the HV-like scheme the vector current
is conserved.\\
\\
In the four-dimensional scheme we find however
\bq
A^{VVA}_{FD} & = & \int \frac{d^dk}{(2\pi)^d i} \frac{\left(k^{(m_\eps)}\right)^2}{k_0^2 k_1^2 k_2^2} 
8 i \eps_{\beta\mu\lambda\kappa} k_1^\lambda p_1^\kappa \nonumber \\
& = & \frac{1}{3} \frac{1}{(4\pi)^2} 8 i \eps_{\beta\mu\lambda\kappa} p_1^\lambda p_2^\kappa
\eq
and the vector current is not conserved.

\subsection{The AAA-anomaly}

The AAA-anomaly is obtained by replacing the vertices $\gamma_\alpha$ and $\gamma_\beta$ in fig. (\ref{singlet})
by $\gamma_\alpha \gamma_5$ and $\gamma_\beta \gamma_5$, respectively.
In the 't Hooft-Veltman scheme we now have to evaluate terms like
\bq
2 \mbox{Tr}\; k\!\!\!/_2 \gamma_\alpha \gamma_5 k\!\!\!/_0 \gamma_\beta \gamma_5 k\!\!\!/_1 k\!\!\!/^{(m_\eps)} \gamma_5
& = & 
2 \mbox{Tr} \; k\!\!\!/_2 \gamma_\alpha k\!\!\!/_0 \gamma_\beta k\!\!\!/_1 k\!\!\!/^{(m_\eps)} \gamma_5
+ 4 \left(k_{(m^\eps)}\right)^2 \mbox{Tr}\; \gamma_\alpha \gamma_\beta k\!\!\!/_1 k\!\!\!/_2 \gamma_5 \nonumber \\
& = & 2 \left(k^{(m_\eps)}\right)^2 \eps(\alpha,\beta,p_1,p_2) + 4 \left(k^{(m_\eps)}\right)^2
\eps(\alpha,\beta,k,p_1+p_2). \nonumber \\
\eq
For the anomaly we then obtain
\bq
A_{HV}^{AAA} & = & \int \frac{d^dk}{(2 \pi)^d i} \frac{\left(k^{(m_\eps)}\right)^2}{k_0^2 k_1^2 k_2^2}
\left( 4 \eps(\alpha,\beta,p_1,p_2) + 8 \eps(\alpha,\beta,k,p_1+p_2) \right) \nonumber \\
& = & \frac{1}{3} \frac{1}{(4 \pi)^2} 8 i \eps_{\alpha\beta\lambda\kappa} p_1^\lambda p_2^\kappa. 
\eq
The calculation is simpler in the four-dimensional scheme. Since all Dirac matrices are in four dimensions, we are allowed
to permute the two additional $\gamma_5$'s next to each other and we obtain the same result as for the AVV-anomaly:
\bq
A_{FD}^{AAA} & = & \frac{1}{3} \frac{1}{(4 \pi)^2} 8 i \eps_{\alpha\beta\lambda\kappa} p_1^\lambda p_2^\kappa
\eq

\subsection{The non-singlet axial-vector current}

The Ward identity for the non-singlet axial-vector current for massless fermions reads
\bq
\label{ward}
( p_{1} - p_{2} )^{\mu} \Gamma_{\mu 5} & = & {S}_{F}^{-1}(p_{1}) \gamma_{5} + \gamma_{5} {S}_{F}^{-1}(p_{2}),
\eq
where $i S_F(p)$ denotes the full fermion propagator and $\Gamma_{\mu 5}$ denotes the full 
$\gamma_\mu \gamma_5$-vertex. We are now going to check the Ward identity at one-loop level.
\begin{figure}
\begin{center}
\begin{picture}(300,100)(0,0)
\ArrowLine(50,50)(90,70)
\ArrowLine(90,30)(50,50)
\Vertex(50,50){2}
\Photon(10,50)(50,50){4}{3}
\Gluon(80,65)(80,35){4}{2}
\ArrowLine(150,50)(190,70)
\ArrowLine(190,30)(150,50)
\Vertex(150,50){2}
\Photon(110,50)(150,50){4}{3}
\GlueArc(170,60)(10,210,30){4}{4}
\ArrowLine(250,50)(290,70)
\ArrowLine(290,30)(250,50)
\Vertex(250,50){2}
\Photon(210,50)(250,50){4}{3}
\GlueArc(270,40)(10,330,150){4}{4}
\Text(0,60)[lb]{$p_1-p_2$}
\Text(90,80)[b]{$p_1$}
\Text(90,20)[t]{$p_2$}
\end{picture}
\caption{\label{nonsinglet} Feynman graphs for the non-singlet axial-vector Ward identity}
\end{center}
\end{figure}
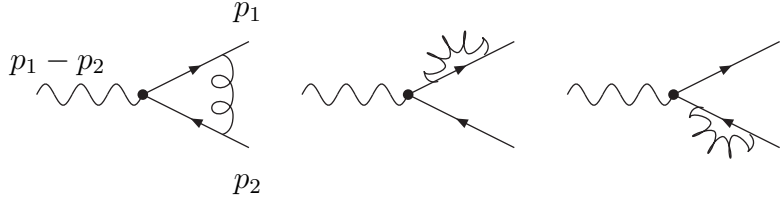
The relevant diagrams are shown in fig.\ref{nonsinglet}.
The momentum $p_1$ is flowing outwards, whereas we take the momentum $p_2$ to be directed inwards.
We start with the calculation in the four-dimensional scheme.
The one-loop contribution from the right-hand-side of eq. (\ref{ward}) reads:
\bq
- \int \frac{d^dk}{(2 \pi)^d i} \frac{\gamma_\nu k\!\!\!/_2 \gamma^\nu}{k_0^2 k_2^2} \gamma_5
- \gamma_5 \int \frac{d^dk}{(2 \pi)^d i} \frac{\gamma_\nu k\!\!\!/_1 \gamma^\nu}{k_0^2 k_1^2} \gamma_5,
\eq
where we used the notation $k_0 = k$, $k_1 = k+p_2$ and $k_2 = k + p_1$.
The contribution from the three-point diagram reads:
\bq
\int \frac{d^dk}{(2 \pi)^d i} \frac{\gamma_\nu k\!\!\!/_2 \gamma_\mu \gamma_5 k\!\!\!/_1 \gamma^\nu}{k_0^2 k_1^2 k_2^2}
\eq
Contracting with $(p_1-p_2)^\mu$ and rewriting $p_1-p_2=k_2-k_1$ we obtain
\bq
k\!\!\!/_2 \left( p\!\!\!/_1 - p\!\!\!/_2 \right) \gamma_5 k\!\!\!/_1 & = & 
k\!\!\!/_2 \left( k\!\!\!/_2 - k\!\!\!/_1 \right) \gamma_5 k\!\!\!/_1 \nonumber \\
&= & k_2^2 \gamma_5 k\!\!\!/_1 + k_1^2 k\!\!\!/_2 \gamma_5  - \left( k^{(m_\eps)} \right)^2 \left( p\!\!\!/_1 - 
p\!\!\!/_2 \right) \gamma_5.
\eq 
The first two terms correspond exactly to the right-hand-side of eq.(\ref{ward}). However, the third term
in the equation above does not vanish. This term yields
\bq
 \frac{1}{(4 \pi)^2} (p\!\!\!/_1 - p\!\!\!/_2) \gamma_5.
\eq
In order to restore the Ward identity we have to perform a finite renormalization on the 
non-singlet axial-vector current
\bq
\Gamma_{\mu5}^r & = & Z_{5,FD}^{ns} \Gamma_{\mu5}^0,
\eq
where $Z_{5,FD}^{ns}$ is given by (including a factor $g^2 C_F$, where 
$ C_F = \mbox{Tr} T^a T^a$ is the fundamental Casimir of the gauge group)
\bq
Z_{5}^{ns} & = & 1 + \frac{\alpha_s}{4 \pi} C_F,
\eq
with $\alpha_s = g^2 /(4 \pi)$.
In the 't Hooft-Veltman scheme we have to replace the vertex $\gamma_\mu \gamma_5$ by the hermitian
expression
\bq
\frac{1}{2} \left( \gamma_\mu \gamma_5 - \gamma_5 \gamma_\mu \right).
\eq
We then obtain
\bq
\lefteqn{ \frac{1}{2} \left( \gamma_\nu k\!\!\!/_2 \left( p\!\!\!/_1 - p\!\!\!/_2 \right) \gamma_5 k\!\!\!/_1 \gamma^\nu
-\gamma_\nu k\!\!\!/_2 \gamma_5 \left( p\!\!\!/_1 - p\!\!\!/_2 \right) k\!\!\!/_1 \gamma^\nu \right) = } \nonumber \\
& = &- k_1^2 \gamma_\nu k\!\!\!/_2 \gamma^\nu \gamma_5 - k_2^2 \gamma_5 \gamma_\nu k\!\!\!/_1 \gamma^\nu \nonumber \\
& & + 4 \eps k_1^2 k\!\!\!/_2 \gamma_5 + 4 \eps k_2^2 \gamma_5 k\!\!\!/_1 
- \left( k^{(m_\eps)} \right)^2  \gamma_\nu \left( \gamma_5 k\!\!\!/_1 + k\!\!\!/_2 \gamma_5 \right) \gamma^\nu.
\eq
The last line spoils the Ward identity. This line gives the contribution
\bq
- 4 \frac{1}{(4 \pi)^2} \left( p\!\!\!/_1 - p\!\!\!/_2 \right) \gamma_5.
\eq
After inclusion of the overall factor $g^2 C_F$
the finite renormalization constant in the 't Hooft-Veltman scheme is given by
\bq
Z_{5,HV}^{ns} & = & 1 -4 \frac{\alpha_s}{4 \pi} C_F
\eq
in agreement with the literature \cite{Larin}.

\subsection{The Ward identity for the vector current}

The Ward identity for the vector current reads
\bq
\label{ward_vector}
( p_{1} - p_{2} )^{\mu} \Gamma_{\mu} & = & {S}_{F}^{-1}(p_{1}) - {S}_{F}^{-1}(p_{2}).
\eq
The contribution from the three-point integral, contracted with $(p_1-p_2)^\mu$ reads:
\bq
\int \frac{d^dk}{(2 \pi)^d i} \frac{\gamma_\nu k\!\!\!/_2 \left( p\!\!\!/_1 -p\!\!\!/_2 \right) k\!\!\!/_1 \gamma^\nu}{k_0^2 k_1^2 k_2^2}
\eq
In the 't Hooft-Veltman scheme we are allowed to write
\bq
\gamma_\nu k\!\!\!/_2 \left( p\!\!\!/_1 -p\!\!\!/_2 \right) k\!\!\!/_1 \gamma^\nu & = & \gamma_\nu k\!\!\!/_2 \left( k\!\!\!/_2 -k\!\!\!/_1 \right) k\!\!\!/_1 \gamma^\nu
\nonumber \\
& = & k_2^2 \gamma_\nu k\!\!\!/_1 \gamma^\nu - k_1^2 \gamma_\nu k\!\!\!/_2 \gamma^\nu
\eq
and the Ward identity is satisfied.
In the four-dimensional scheme we find
\bq
\gamma_\nu k\!\!\!/_2 \left( p\!\!\!/_1 -p\!\!\!/_2 \right) k\!\!\!/_1 \gamma^\nu & = & \left(k_2^{(2)} \right)^2 \gamma_\nu k\!\!\!/_1 \gamma^\nu - \left( k_1^{(2)} \right)^2 \gamma_\nu k\!\!\!/_2 \gamma^\nu \nonumber \\
& = & k_2^2 \gamma_\nu k\!\!\!/_1 \gamma^\nu - k_1^2 \gamma_\nu k\!\!\!/_2 \gamma^\nu
-2  \left( k^{(m_\eps)} \right)^2 \left( p\!\!\!/_1 - p\!\!\!/_2 \right).
\eq
The Ward identity is violated by the last term. This term gives a contribution
\bq
-2 ( p\!\!\!/_1 - p\!\!\!/_2 ) \int \frac{d^dk}{(2 \pi)^d i} \frac{\left(k^{(m_\eps)}\right)^2}{k_0^2 k_1^2 k_2^2}
& = & \frac{1}{(4 \pi)^2} ( p\!\!\!/_1 - p\!\!\!/_2 )
\eq
to the left-hand side of eq. (\ref{ward_vector}). The appropriate finite renormalization constant is therefore
(again with the factor $g^2 C_F$ included):
\bq
Z^{ns}_{FD} & = & 1 + \frac{\alpha_s}{4 \pi} C_F.
\eq
Within the original formulation of dimensional reduction one could argue that the four-dimensional
$(k_1^{(2)})^2$ and $(k_2^{(2)})^2$ cancel exactly the $d$-dimensional propagators
$k_1^2$ and $k_2^2$, since $k_1^2$ has ``less components'' than $(k_1^{(2)})^2$. (We have $d<4$.)
One would then conclude that the Ward identity for the vector current is not violated.
However, as already mentioned in the introduction, this scheme is inconsistent:
In that case one would also have a projection
onto the $d$-dimensional subspace and one has to conclude that the trace of four Dirac
matrices with one $\gamma_5$ vansihes in the regularized theory.
In our approach we avoid this inconsistency by distinguishing the rank (which may be smaller than 4) from
the dimension of the representative (which will always be larger than 4).
Therefore the propagators are not cancelled exactly and we need a finite renormalization of the vector current.

\subsection{Summary of the examples}

In the previous section we have computed various triangle anomalies in both the HV-like and the four-dimensional scheme.
We considered the anomalous divergence of the axial current, when the two other currents in the triangle
were vector currents (AVV anomaly) or axial currents (AAA anomaly).
We also calculated the divergence of the vector current, when one of the remaining two currents was
a vector current and the other one an axial current (VVA anomaly).
We have summarized the results in table \ref{sum_anomaly}.
\begin{table}
\begin{center}
\begin{tabular}{|c|c|c|c|}
\hline 
& AVV & VVA & AAA \\ \hline \hline 
& & & \\
HV & $ 1 $ & $ 0 $ & $ 1/3 $ \\ & & & \\ \hline & & & \\ 
FD & $ 1/3 $ & $ 1/3 $ & $ 1/3 $  \\ & & & \\ \hline 
\end{tabular}
\end{center}
\caption{\label{sum_anomaly} The AVV-, VVA-  and AAA-triangle anomalies in terms of
$8i \eps_{\alpha\beta\lambda\kappa}p_1^\lambda p_2^\kappa / (4\pi)^2$ 
in the HV-like and in the four-dimensional scheme.}
\end{table}
We would like to point out that these results are consistent with the Bardeen relations \cite{Bardeen}:
If the vector current is conserved, the anomalous divergence of the vector and the axial current is according to Bardeen given by
\bq
\label{Bardeen_eq}
\partial^\mu J_\mu^a & = & 0 ,
\nonumber \\
\partial^\mu J_{5\mu}^a & = & 
\frac{1}{(4\pi)^2} \eps_{\mu\nu\sigma\tau} \mbox{Tr} \; \left\{ T_A^a \left(
F_V^{\mu\nu} F_V^{\sigma\tau} + \frac{1}{3} F_A^{\mu\nu} F_A^{\sigma\tau}
+ \frac{i}{6} A^\mu A^\nu F_V^{\sigma\tau} 
+ \frac{i}{6} F_V^{\mu\nu} A^\sigma A^\tau \right. \right. \nonumber \\
& & \left. \left. + \frac{i}{6} A^\mu F_V^{\nu\sigma} A^\tau
-\frac{2}{3} A^\mu A^\nu A^\sigma A^\tau \right) \right\},
\eq
where
\bq
V^\mu & = & T_V^a V_a^\mu, \nonumber \\
A^\mu & = & T^a_A A_a^\mu, \nonumber \\
F_V^{\mu\nu} & = & \partial^\mu V^\nu - \partial^\nu V^\mu - i \left[ V^\mu, V^\nu \right] - i \left[ A^\mu, A^\nu \right], \nonumber \\
F_A^{\mu\nu} & = & \partial^\mu A^\nu - \partial^\nu A^\mu - i \left[ V^\mu, A^\nu \right] - i \left[ A^\mu, V^\nu \right] 
\eq
and the trace is over the internal degrees of freedom (such as weak isospin for example). 
Equation (\ref{Bardeen_eq}) corresponds to the HV-like scheme.
The four-dimensional scheme treats the vector and axial current symmetrically. In this case the 
anomalous divergences of the currents are according to Bardeen given by
\bq
\partial^\mu J_\mu^a & = & 
\frac{1}{(4\pi)^2} \eps_{\mu\nu\sigma\tau} \mbox{Tr} \; \left\{ T_A^a \gamma_5 \left[
\frac{1}{3} \partial^\mu W^\nu \partial ^\sigma W^\tau \right. \right. \nonumber \\
& & \left. \left. - \frac{i}{6} \left( 
 \partial^\mu W^\nu W^\sigma W^\tau - W^\mu \partial^\nu W^\sigma W^\tau + W^\mu W^\nu \partial^\sigma W^\tau \right)
\right] \right\}, \nonumber \\
\partial^\mu J_{5\mu}^a & = & 
\frac{1}{(4\pi)^2} \eps_{\mu\nu\sigma\tau} \mbox{Tr} \; \left\{ T_A^a \left[
\frac{1}{3} \partial^\mu W^\nu \partial ^\sigma W^\tau \right. \right. \nonumber \\ 
& & \left. \left. - \frac{i}{6} \left( 
 \partial^\mu W^\nu W^\sigma W^\tau - W^\mu \partial^\nu W^\sigma W^\tau + W^\mu W^\nu \partial^\sigma W^\tau \right)
\right] \right\} \nonumber \\
\eq
with
\bq
W^\mu & = & V^\mu + A^\mu \gamma_5
\eq
and the trace is over the internal degrees of freedom (such as weak isospin) as well as over the Dirac matrices.\\
\\
We have also considered open fermion lines and have calculated at one loop the finite renormalization
constants needed to restore the corresponding Ward identities. The results are summarized in 
table \ref{fin_renorm}. 
\begin{table}
\begin{center}
\begin{tabular}{|c|c|c|}
\hline 
& V & A \\ \hline \hline 
& & \\
HV & $ 1 $ & $ 1-4 \frac{\alpha_s C_F}{4\pi} $ \\ & & \\ \hline & & \\ 
FD & $ 1+ \frac{\alpha_s C_F}{4\pi} $ & $ 1+ \frac{\alpha_s C_F}{4\pi} $  \\ & & \\\hline 
\end{tabular}
\end{center}
\caption{\label{fin_renorm} The finite renormalizations needed for the non-singlet axial current (A) and the vector current (V) in the HV-like and in the four-dimensional scheme.}
\end{table}
The results in the HV-like scheme agree with the literature \cite{Larin}.
In the HV-like scheme only the $\gamma_\mu \gamma_5$-vertex needs a finite renormalization, whereas the
$\gamma_\mu$-vertex does not.
In the four-dimensional scheme both the $\gamma_\mu \gamma_5$-vertex and the $\gamma_\mu$-vertex need a finite
renormalization.

\section{Conclusions}

In this paper we have developed a new approach to dimensional regularization.
Instead of the traditional approach, which uses infinite-dimensional vector spaces in order to define
an integration in non-integer dimension, we used K-theory and worked with finite-dimensional vectorspaces only.
We distinguished between the rank of an object and the dimension of a representative of that object.
$d$-dimensional integration corresponds in our framework to integration over an object with rank $d$.
If $d=p+iq$ where $p,q \in {\Bbb Q}$ we were able to construct finite-dimensional vector spaces,
which represent an object of rank $p+iq$.
Since the set of all $p+iq$ with $p,q \in {\Bbb Q}$ is a dense subset of the complex $d$-plane, we were able to extend
the construction by a limiting procedure to the whole complex $d$-plane.
It is possible to work with vectorspaces of even dimension only.
We then considered the continuation of the Dirac algebra.
We defined two schemes, one similar to the 't Hooft-Veltman scheme, the other one similar to the
four-dimensional helicity scheme.
Although we maintained at each step Lorentz invariance (also for the HV-like scheme), it is not possible
to preserve at the same time the structure of the Clifford algebra. We showed that the two schemes
correspond to two different deformations of the Clifford algebra.
The HV-like scheme deforms
\bq
\left\{ \gamma^\mu, \gamma_5 \right\} & = & 0,
\eq
whereas the four-dimensional scheme deforms
\bq
\left\{ \gamma^\mu, \gamma^\nu \right\} & = & 2 g^{\mu\nu} 1.
\eq
In practical calculations the HV-like scheme behaves like the original scheme of 't Hooft and Veltman.
The main result of this paper is the consistent definition of the four-dimensional scheme.
It is the purpose of this paper to advocate the four-dimensional scheme for future calculations.
There are a few points, where the definition of the four-dimensional scheme given here differs from the 
definition of the four-dimensional helicity scheme \cite{FDH}.
We summarize them here:
\begin{itemize}
\item Our regularization scheme is defined also for parity-violating
amplitudes.
\item In our scheme there is for
\bq
\mbox{rank}(X_0 + X_1) < 4
\eq
no projection from $X_0$ onto $X_0+X_1$.
\item In our scheme there is for
\bq
\mbox{rank}(X_0+X_1) < 4
\eq
always a projection from $X_0+X_1$ to $X_0$.
\item In our scheme we need a finite renormalization of the vector current 
and the axial vector current
in order to restore the relevant Ward identities.
\end{itemize}
Our scheme, like dimensional reduction or like the original formulation 
of the four-dimensional helicity scheme, respects supersymmetry and is therefore
well suited to regulate supersymmetric theories. It is free of inconsistencies inherent in the latter
two schemes.
It is therefore a candidate for a consistent regularization scheme for supersymmetric theories.\\
\\
Finally we would like to remark that in order to perform a calculation entirely in our scheme, one needs
usually besides the
loop amplitudes also splitting functions and anomalous dimensions.
Consistency requires that all quantities are calculated in the same scheme. 

\section{Acknowledgements}

I would like to thank Michel Bauer, Sergei Larin, Sven Moch, Henri Navelet, Carlos Savoy, Peter Uwer and Andrew Waldron
for useful discussions.
Many thanks to Eric Laenen for his comments on the manuscript.

\begin{appendix}

\section{Notation}

Here we would like to explain the notation, which is used throughout the paper.
Let $X_1$ and $X_2$ be vectorspaces of dimension $2 m_1$ and $2 m_2$, respectively.
The direct sum $X_1 + X_2$ is then a vectorspace of dimension $2(m_1+m_2)$.
We may identify $X_1$ and $X_2$ with subspaces of $(X_1+X_2)$ in the standard way. 
We use the notation
\bq
x_{(m_1)}^\mu
\eq
to denote a vector of $(X_1+X_2)$, which lies entirely in the $X_1$-subspace of $(X_1+X_2)$.
If we choose a specific coordinate system of $(X_1+X_2)$, such that the first $2m_1$ coordinates
refer to $X_1$, the remaining $2m_2$ coordinates to $X_2$, only the first $(2m_1)$ components of
$x_{(m_1)}^\mu$ are non-zero. \\
\\
Contractions with Dirac matirces and scalar products are denoted as follows:
\bq
p\!\!\!/_{(m)} & = & \sum\limits_{\mu=0}^{2m-1} p^{(m)}_\mu \gamma^\mu_{(m)}, \\
(p q)_{(m)} & = & 2 \sum\limits_{\mu=0}^{2m-1} p^{(m)}_\mu q_{(m)}^\mu .
\eq
Next we consider a vector bundle over $X_1$, whose fibre is isomorph to a vectorspace spanned by the
basis vectors $T^a$.
In order to have a concrete example we may take the $T^a$'s to be the generators of a Lie group
$SU(N)$. The index $a$ runs then from 1 to $N^2-1$.
We may consider a similar construction over $X_2$. In order to distinguish the fibres over
$X_1$ and $X_2$, we denote the basis vectors by $T^a_{(m_1)}$ and $T^a_{(m_2)}$, respectively.
In the example above, the index $a$ runs in both cases from $1$ to $N^2-1$. The generators
satisfy the standard Lie algebra relations
\bq
\left[ T^a_{(m_1)}, T^b_{(m_1)} \right] & = & i f^{abc}_{(m_1)} T^c_{(m_1)}, \nonumber \\
\left[ T^a_{(m_2)}, T^b_{(m_2)} \right] & = & i f^{abc}_{(m_2)} T^c_{(m_2)} ,
\eq
but since $T^a_{(m_1)}$ and $T^a_{(m_2)}$ generate two different copies of $SU(N)$, they commute
with each other:
\bq
\left[ T^a_{(m_1)}, T^b_{(m_2)} \right] & = & 0 .
\eq

\section{Fibre bundle construction for the gauge groups}

In this appendix we reformulate the different prescriptions of section 5 for the gauge group in a
more mathematical language.
Let $E_1 \stackrel{\pi_1}{\rightarrow} X_1$ and $E_2 \stackrel{\pi_2}{\rightarrow} X_2$
be vectorbundles with base spaces $X_1$ and $X_2$, total spaces $E_1$ and $E_2$ and projections
$\pi_1$ and $\pi_2$. We assume that the base spaces have dimensions
$2m_1$ and $2m_2$ and denote the dimensions of the fibres by $n_1$ and $n_2$. We denote
the structure groups by $G_1$ and $G_2$, respectively.
Then we form the direct sum $E_1 + E_2$. This is a vectorspace of dimension $(2m_1 + n_1) + (2m_2 + n_2)$.
We have a natural $(G_1 \times G_2)$-action on $E_1+E_2$, induced by the action of $G_1$ on $E_1$ and by
the action of $G_2$ on $E_2$.\\
Now we may specify $G_1=G$ and $G_2=\{e\}$, where $G$ is the gauge group ($SU(3)$ for example). This corresponds
to prescription 3 in section 5.\\
\\
The choice $G_1=G$ and $G_2=G$ leads to a $(G \times G)$-action on $E_1+E_2$ (e.g. two different copies of $G$ are 
acting on $E_1+E_2$) and corresponds to prescription 2.\\
\\
Using the diagonal map $\Delta : G \rightarrow G \times G$ given by $g \rightarrow (g,g)$ and the choice $G_1=G$ and $G_2=G$
as above, we obtain a $G$-action on $E_1+E_2$ (e.g. only one copy of $G$ acts on both $E_1$ and $E_2$).
This corresponds to prescription 1.

\section{Integration on a tensor product}

In section 3 we have viewed $X_1 \cdot X_2$ as vector space of dimension $4 m_1 m_2$ and have obtained for the integration
over $X_1 \cdot X_2$:
\bq
\int\limits_{X_1 \cdot X_2} d^{d_1 d_2} k f(k^2) & = & 
\frac{\pi^{d_1d_2/2}}{\Gamma\left(\frac{d_1d_2}{2}\right)} \int dk^2 \left( k^2 \right)^{d_1d_2/2-1} f(k^2)
\eq
On the other hand we may view $X_1 \cdot X_2$ embedded in a trivial vector bundle with base $X_1$
and fibre $X_1 \cdot X_2$. We then may decide to integrate over all fibres first and in the end integrate
over the base space. We do this as follows: We introduce coordinates $z_{a b}$ on $X_1 \cdot X_2$,
with $0 \le a \le 2 m_1-1$ and $0 \le b \le 2 m_2-1$ and coordinates $x_{c}$ on $X_1$
with $0 \le c \le 2 m_1 -1$.
We then obtain
\bq
\int\limits_{X_1 \cdot X_2} d^{d_1 d_2} k f(k^2) & = &
\frac{\pi^{d_1 d_2/2-2m_1m_2}}{\Gamma\left(\frac{d_1d_2}{2}-2m_1 m_2\right)} \int d z_{a b} \int dk_\perp^2 
\left( k_\perp^2 \right)^{d_1 d_2/2-2m_1 m_2 -1}
f \left( \sum\limits_{a,b} z_{a b}^2 +k_\perp^2\right) \nonumber \\
& = & \frac{\pi^{d_1 d_2/2-2m_1m_2}}{\Gamma\left(\frac{d_1d_2}{2}-2m_1 m_2\right)} 
\int dx_{a} \int d z_{a b} \int dk_\perp^2 
\left( k_\perp^2 \right)^{d_1 d_2/2-2m_1 m_2 -1} \nonumber \\
& & \cdot f \left( \sum\limits_{a,b} z_{a b}^2 + k_\perp^2 \right) 
\prod\limits_{a}
\sqrt{\sum\limits_{b} z_{a b}^2} \;
\delta\left( \sum\limits_{b} z_{a b}^2 - x_{a}^2 \right)
\nonumber \\
& = & \frac{\pi^{d_1 d_2/2-2m_1m_2}}{\Gamma\left(\frac{d_1d_2}{2}-2m_1 m_2\right) \Gamma(m_2)^{2 m_1}} \int dx_{c} 
\int dk_\perp^2 
\left( k_\perp^2 \right)^{d_1 d_2/2-2m_1 m_2 -1} \nonumber \\
& & \cdot \left( \prod\limits_{c} \left| x_{c} \right|^{2 m_2-1} \right) 
f \left( \sum\limits_{c} x_{c}^2 + k_\perp^2 \right).
\eq
In the second line we introduced first a delta-function and performed then the integration
over the $z_{ab}$-variables. We have therefore performed the integration over the fibre $X_1 \cdot X_2$.
We want to check that the remaining integration over $X_1$ gives us back the result of eq. (\ref{tensor}).
We introduce spherical coordinates on $X_1$:
\bq
x_0 & = & r \cos \theta_1, \nonumber \\
x_1 & = & r \sin \theta_1 \cos \theta_2, \nonumber \\
& ... & \nonumber \\
x_{2m_1-2} & = & r \cos \theta_1 \cos \theta_2 ... \cos \theta_{2m_1-1}, \nonumber \\
x_{2m_1-1} & = & r \cos \theta_1 \cos \theta_2 ... \sin \theta_{2m_1-1}.
\eq
The integral becomes then
\bq
\int\limits_{X_1 \cdot X_2} d^{d_1 d_2} k f(k^2) & = & 
\frac{\pi^{d_1d_2/2}}{\Gamma\left(\frac{d_1d_2}{2}-2 m_1m_2\right) \Gamma(m_2)^{2m_1}}
\int dk_\perp^2 
\left( k_\perp^2 \right)^{d_1 d_2/2-2m_1 m_2 -1} \nonumber \\
& & \cdot \int\limits_0^\infty dr \; r^{4 m_1 m_2 -1} f(r^2+k_\perp^2) \nonumber \\
& & \cdot \left(\prod\limits_{l=1}^{2m_1-2} \int\limits_0^\pi d\theta_l 
 \left( \sin \theta_l \right)^{2m_1-1-l} 
 \left| \cos \theta_l  \left( \sin \theta_l \right)^{2m_1-l} \right|^{2m_2-1} \right) \nonumber \\
& & \cdot \int\limits_0^{2 \pi} d\theta_{2m_1-1}
\left| \cos \theta_{2m_1-1}  \sin \theta_{2m_1-1} \right|^{2m_2-1} .
\eq 
The relevant angular integrals are obtained from the following formulae \cite{Erdelyi}: 
\bq
\int\limits_0^{\pi/2} d \theta \left( \sin \theta \right)^{2x-1}
\left( \cos \theta \right)^{2y-1} & = & \frac{1}{2} B(x,y), \nonumber \\
& & \mbox{Re}\;x > 0, \;\;\; \mbox{Re}\; y > 0.
\eq
This formula allows us to evaluate the following integral:
\bq
\int\limits_0^\pi d\theta \left( \sin \theta \right)^{q-1} \left|
\cos \theta \left( \sin \theta\ \right)^{q} \right|^{2p-1} 
& = & B \left( q p,p \right), \nonumber \\
& & \mbox{Re}\;p > 0, \;\;\; \mbox{Re} \; q p > 0.
\eq
From the integral
\bq
\int\limits_0^\pi d\theta  \left( \sin \theta \right)^\alpha \exp \left( i \beta \theta \right) & = &
 \frac{\pi}{2^\alpha} \frac{\Gamma(1+\alpha)}{\Gamma\left( 1 + \frac{\alpha + \beta}{2}\right)
\Gamma\left( 1 + \frac{\alpha - \beta}{2}\right)} \exp\left( \frac{i \pi \beta}{2} \right), \nonumber \\
& & \mbox{Re}\; \alpha > -1, 
\eq
we obtain 
\bq
\int\limits_0^{2\pi} d\theta \left| \cos \theta \sin \theta \right|^{2p-1}
& = & \frac{2^{3-4p} \pi \Gamma(2p)}{\Gamma\left(p+\frac{1}{2}\right)
\Gamma\left(p+\frac{1}{2}\right)}, \nonumber \\
& & \mbox{Re}\;p > 0.
\eq
Performing the angular integrations and noting the identity
\bq
\frac{\Gamma(2p)}{\Gamma(p)\Gamma\left(p+\frac{1}{2}\right)} & = &
\frac{2^{2p-1}}{\sqrt{\pi}}
\eq
we obtain
\bq
\int\limits_{X_1 \cdot X_2} d^{d_1 d_2} k f(k^2) & = & 
\frac{\pi^{d_1d_2/2}}{\Gamma\left(\frac{d_1d_2}{2}-2 m_1m_2\right) \Gamma(2m_1m_2)}
\int dk_\perp^2 
\left( k_\perp^2 \right)^{d_1 d_2/2-2m_1 m_2 -1} \nonumber \\
& & \cdot \int\limits_0^\infty dr^2 \; \left(r^2\right)^{2 m_1 m_2 -1} f(r^2+k_\perp^2) \nonumber \\
& = & \frac{\pi^{d_1d_2/2}}{\Gamma\left(\frac{d_1d_2}{2}\right)} \int dk^2 \left( k^2 \right)^{d_1d_2/2-1} f(k^2)
\eq
in agreement with the result already obtained in eq. (\ref{tensor}).

\section{Integrals}

For the calculation of the anomalies we need the following two integrals:
\bq
\int \frac{d^dk}{(2 \pi)^d i} \frac{\left( k^{(m^\eps)} \right)^2}{k^2 (k+p_1)^2 (k-p_2)^2}
& = & 4 \pi \eps \int \frac{d^{d+2}k}{(2 \pi)^{d+2} i} \frac{\left( k^{(m^\eps)} \right)^2}{k^2 (k+p_1)^2 (k-p_2)^2}
\nonumber \\
& = & - \frac{1}{2} \frac{1}{(4 \pi)^2} + O(\eps)
\eq
The second integral is given by
\bq
\lefteqn{
4 i \eps_{\alpha\beta\lambda\kappa} (p_1+p_2)^\kappa \int \frac{d^dk}{(2 \pi)^d i} \frac{k^\lambda \left( k^{(m^\eps)} \right)^2}{k^2 (k+p_1)^2 (k-p_2)^2}
= } & & \nonumber \\
& = & 4 i \eps_{\alpha\beta\lambda\kappa} (p_1+p_2)^\kappa 
\;2 \int\limits_{0}^1 dx \int\limits_0^{1-x} dy 
\int \frac{d^dk}{(2 \pi)^d i} \frac{-(1-x)p_1^\lambda \left( k^{(m^\eps)} \right)^2}{\left( k^2 + L \right)^3}
\nonumber \\
& = & 4 i \eps_{\alpha\beta\lambda\kappa} p_1^\lambda p_2^\kappa 
\;2 \int\limits_{0}^1 dx (x-1) \int\limits_0^{1-x} dy 
4 \pi \eps \int \frac{d^{d+2}k}{(2 \pi)^{d+2} i} \frac{\left( k^{(m^\eps)} \right)^2}{\left( k^2 + L \right)^3}
\nonumber \\
& = & \frac{1}{3} \frac{1}{(4 \pi)^2} 4 i \eps_{\alpha\beta\lambda\kappa} p_1^\lambda p_2^\kappa + O(\eps),
\eq
where $L$ is given by
\bq
L &= & y p_2^2 + (1-x-y) p_1^2 - \left( (1-x-y)p_1 -y p_2 \right)^2.
\eq

\end{appendix}

\end{document}